\documentclass[prd,twocolumn,preprint numbers, aps,nofootinbib,showpacs]{revtex4-1}
 \usepackage{amsmath}
\usepackage{amssymb}
\usepackage{graphicx}
\usepackage{color}
\usepackage{bm}
\usepackage{times}
\usepackage{hyperref}
\hypersetup{
  colorlinks=true,
  citecolor=blue,
  linkcolor=blue,
  urlcolor=blue}

\usepackage{epsfig}
\usepackage{dcolumn}
\usepackage{float}

\usepackage{epsfig}
\usepackage{dcolumn}
\usepackage[normalem]{ulem}
\def\be{\begin{equation}}
\def\ee{\end{equation}}
\def\bea{\begin{eqnarray}}
\def\eea{\end{eqnarray}}
\def\gsim{\ \rlap{\raise 2pt\hbox{$>$}}{\lower 2pt \hbox{$\sim$}}\ }
\def\lsim{\ \rlap{\raise 2pt\hbox{$<$}}{\lower 2pt \hbox{$\sim$}}\ }
\def\dslash{\kern-4pt \not{\hbox{\kern-2pt $\partial$}}}
\def\pslash{\not{\hbox{\kern-2pt p}}}

\newcommand{\dcp}{\delta_{CP}}
\newcommand{\nova}{NO$\nu$A\ }

\newcommand{\cnv}{\v{C}erenkov}

\begin{document}

\DeclareGraphicsExtensions{.eps,.ps}

\title{Effect of systematics in the T2HK, T2HKK, and DUNE experiments}

\author{Monojit Ghosh}
\email{mghosh@phys.se.tmu.ac.jp}
\affiliation{Department of Physics, Tokyo Metropolitan University, Hachioji, Tokyo 192-0397, Japan}

\author{Osamu Yasuda}
\email{yasuda@phys.se.tmu.ac.jp}
\affiliation{Department of Physics, Tokyo Metropolitan University, Hachioji, Tokyo 192-0397, Japan}

\begin{abstract}

T2HK and T2HKK are the proposed extensions of the of T2K experiments in Japan and DUNE is the future long-baseline program of Fermilab. 
All these three experiments will use extremely high beam power and large detector volumes to observe neutrino oscillation. Because of the large statistics, these experiments will be highly sensitive to
systematics. Thus a small change in the systematics can cause a significant change in their sensitivities. To understand this, we do a comparative study of T2HK, T2HKK and DUNE 
with respect to their systematic errors. Specifically we study the effect of the systematics in the determination of neutrino mass hierarchy, octant of the mixing angle $\theta_{23}$ and $\dcp$ in the standard three
flavor scenario and also analyze the role of systematic uncertainties in constraining the parameters of the nonstandard interactions in neutrino propagation. 
Taking the overall systematics for signal and background normalization, we quantify how the sensitivities of these experiments change if the systematics
are varied from 1\% to 7\%.

\end{abstract}

\maketitle

\section{Introduction}
\label{sec1}
The phenomenon of neutrino oscillation suggests neutrinos have mass and mixing. In standard three flavor, neutrino oscillations are described by three mixing angles i.e., $\theta_{12}$, $\theta_{13}$, $\theta_{23}$,
two mass squared differences i.e., $\Delta m^2_{21}$, $\Delta m^2_{31}$ and one Dirac type CP phase i.e., $\dcp$. Among these six parameters at this moment the unknowns are: 
(i) the neutrino mass hierarchy i.e., normal hierarchy (NH: $\Delta m^2_{31} >0$) or inverted hierarchy (IH: $\Delta m^2_{31} < 0$), 
(ii) the octant of the mixing angle $\theta_{23}$ i.e., lower octant (LO: $\theta_{23} < 45^\circ$)
or higher octant (HO: $\theta_{23} > 45^\circ$) and (iii) the CP phase $\dcp$. As the appearance channel probability, which gives the transition of $\nu_\mu \rightarrow \nu_e$ 
depends on all these three unknowns, in principle, the accelerator based long-baseline experiments can determine them by studying the electron events at the far detector.
Currently running such kind of experiments are T2K \cite{t2k} and \nova \cite{nova}. The recent results of T2K shows a mild preference for normal hierarchy, $\theta_{23}=45^\circ$ and $\dcp=-90^\circ$. 
On the other hand the current
\nova results are in accordance with T2K regarding the fit to the hierarchy and $\dcp$ but it excludes maximal mixing at $2.6 \sigma$ \cite{Adamson:2017qqn}. 
Note that at this moment the statistical significance of these hints are very weak and one needs further data to establish the true nature of these parameters on a firm footing. 

Due to comparatively shorter baselines, low beam power and small detector sizes, the sensitivities of T2K and \nova are limited. 
The shorter baselines of T2K and \nova restrict them to have sensitivity only in the 
favorable parameter space \cite{Prakash:2012az,Agarwalla:2013ju}. 
These experiments also suffer from parameter degeneracy due to less matter effect \cite{Barger:2001yr,Ghosh:2015ena}. Even in the favorable parameter space, these experiments cannot have much
sensitivity for their low statistics \cite{Ghosh:2012px,Ghosh:2013zna,Prakash:2013dua,Ghosh:2014dba,Chatterjee:2013qus,Ghosh:2015tan,Bharti:2016hfb,Soumya:2016aif}. 
Thus it is the job of the future high statistics long-baseline experiments to determine the remaining unknowns in the neutrino oscillation in a conclusive manner.
The example of such experiments are T2HK \cite{Abe:2014oxa}, T2HKK \cite{T2HKK} and DUNE \cite{Acciarri:2015uup}. 
Due to smaller baseline T2HK will not have much 
hierarchy sensitivity in the unfavorable parameter space but it can have excellent sensitivity in the favorable parameter space \cite{Fukasawa:2016yue,Ballett:2016daj}. Thus if nature
chooses a favorable value of the unknown parameters then T2HK will be sufficient to determine them at a conclusive level. 
Apart from establishing the true nature of the unknown parameters in the standard three neutrino picture, these future long-baseline experiments can also probe different new physics scenarios like 
nonstandard interactions (NSI) in neutrino propagation \cite{Wolfenstein:1977ue,Guzzo:1991hi,Roulet:1991sm,Ohlsson:2012kf,Miranda:2015dra}.
NSI has caught a lot of attention particularly
because Ref.\,\cite{Gonzalez-Garcia:2013usa} pointed out that there is a
tension between the mass-squared difference deduced from the
solar neutrino observations and the one from the KamLAND experiment, and that the tension can be
resolved by introducing the flavor-dependent NSI
in neutrino propagation.
Recent studies of NSI in neutrino propagation for the long-baseline experiments can be found in 
 \cite{Friedland:2012tq,Adhikari:2012vc,Masud:2015xva,deGouvea:2015ndi,Rahman:2015vqa,Coloma:2015kiu,Liao:2016hsa,Soumya:2016enw,Blennow:2016etl,Forero:2016cmb,Huitu:2016bmb,Bakhti:2016prn,
 Masud:2016bvp,Coloma:2016gei,Masud:2016gcl,Agarwalla:2016fkh,Ge:2016dlx,Liao:2016bgf,Fukasawa:2016gvm,Blennow:2016jkn,Liao:2016orc,Deepthi:2016erc,Fukasawa:2016lew}. 
All these experiments mentioned above will use high beam power and large
detectors. Thus it is easy to understand that these experiments will be highly sensitive to the effect of the systematics. 
The systematic errors in the long-baseline experiments arise mainly
from the uncertainties related to fluxes and cross sections.
Studies of different sources of systematic uncertainties affecting the measurement of neutrino oscillation parameters can be found in 
Refs. \cite{Huber:2002mx,Ohlsson:2003ip,Barger:2007jq,Huber:2007em,Coloma:2011pg,Tang:2009na,Ankowski:2016jdd}.
Due to large number of event sample at the far detector, a slight improvement in these
systematic errors can improve sensitivity of these experiments significantly. Thus it will be quite intriguing to see how the sensitivity of these experiments to determine hierarchy, octant and CP violation 
in the standard three flavor scenario as well as to constrain the NSI parameters assuming the existence of NSI in nature
depends on the systematic error. Adopting a very simplistic treatment \footnote{A detailed analysis of the systematics in long-baseline experiments can be found in \cite{Coloma:2012ji}.}, 
we express the systematic errors in terms of an overall signal and background normalization and present our results as function of the systematics
for T2HK, T2HKK and DUNE.

The plan of the paper goes as follows. In the next section we will discuss the experimental specification and simulation details. We will also mention our treatment of the systematics in that section.
In Section \ref{sec3} we will present the hierarchy, octant and CP violation sensitivity of as a function of the systematic error in the standard three flavor scenario. In Section \ref{sec4}, we give
our results corresponding to nonstandard interactions for different values of the systematic errors.
Finally in Section \ref{sec5} we will 
summarize our results and give our conclusions.

\section{Experimental and Simulation Details}
\label{sec2}

For our analysis we assume that under the T2HK project the two water \cnv\ detector tanks each of fiducial mass 187 kt will be placed at the Kamioka site and for the T2HKK experiment
one of the tanks will be at the Kamioka and the other in Korea. The neutrino source for both the setups is J-PARC and the baseline lengths are 295 km for Kamioka and 1100 km for Korea. 
Depending on the locations in Korea, the beam from J-PARC will reach the detector at different off-axis angles. For the present work we consider the flux options of $2.5^\circ$, $2.0^\circ$ and $1.5^\circ$.
The off-axis angle for the Kamioka site is $2.5^\circ$. We have considered a beam power of 1.3 MW with a total exposure of $27 \times 10^{21}$ protons on target (pot). This corresponds to a 10 year running
of both the experiments. Following the T2HKK report, we have divided this runtime in 1:3 for neutrino and antineutrino mode to compensate the lower antineutrino cross sections \cite{T2HKK}. 
We have matched our events with the latest T2HKK report and our results are consistent with their
sensitivity \cite{T2HKK}. For DUNE we use a beam power of 1.2 MW leading to a total exposure of $10 \times 10^{21}$ pot. We assume a liquid argon detector of fiducial mass 40 kt. The baseline for DUNE is 1300 km.
We take the ratio to neutrino and antineutrino run 
as 1:1. Our simulation results of DUNE is consistent with \cite{Acciarri:2015uup}.
We have used the GLoBES \cite{Huber:2004ka,Huber:2007ji} and MonteCUBES \cite{Blennow:2009pk} softwares  to simulate all the above experiments.

We estimate the sensitivity of each experiment in terms of $\chi^2$. 
We calculate the statistical $\chi^2$ by comparing the true events $N^{{\rm true}}$ and test events $N^{{\rm test}}$ using the following Poisson formula:
\begin{eqnarray}
 \chi^2_{{\rm stat}} = \sum_i 2 \bigg[ N^{{\rm test}}_i - N^{{\rm true}}_i - N^{{\rm true}}_i \log\bigg(\frac{N^{{\rm test}}_i}{N^{{\rm true}}_i}\bigg) \bigg]
\end{eqnarray}
where the index $i$ corresponds to the number of energy bins. To incorporate the effect of the systematics, we deviate the test events by
\begin{eqnarray}
 N^{{\rm test}}_i \rightarrow N^{{\rm test}}_i \bigg(1+ \sum_k c_i^k \xi_k \bigg)
\end{eqnarray}
where $c_i^k$ is the $1 \sigma$ systematic error corresponding to the pull variable $\xi_k$. Here the index $k$ stands for number of pull variables. After modifying the events the 
combined statistical and systematic $\chi^2$ is calculated as
\begin{eqnarray}
 \chi^2_{{\rm stat+sys}} = \chi^2_{{\rm stat}} + \sum_k \xi_k^2
\end{eqnarray}
The final $\chi^2$ is obtained by varying $\xi_k$ from $-3$ to $+3$ corresponding to their $3 \sigma$ ranges and minimizing over $\xi_k$ i.e.,
\begin{eqnarray}
 \chi^2 = {\rm min}\{\xi_k\}\big[\chi^2_{{\rm stat+sys}} \big]
\end{eqnarray}
For our analysis of the long-baseline experiments we take four pull variables. These variables are: (i) signal normalization error, (ii) signal tilt error, (iii) background normalization error
and (iv) background tilt error. The normalization error affect the scaling of the events whereas the tilt error or the energy calibration error affects the energy dependence of the events.
The tilt error is incorporated in our analysis by varying the test events in the following way:
\begin{eqnarray}
 N^{{\rm test}}_i \rightarrow N^{{\rm test}}_i \bigg(1+ \sum_k c_i^k \xi_k \frac{E_i - E_{\rm av}}{E_{\rm max} - E_{\rm min}} \bigg)
\end{eqnarray}
where $E_{i}$ is the energy in the $i^{\rm th}$ bin, $E_{\rm min}$ is the lower limit of the full energy range, $E_{\rm max}$ is the higher limit of the full energy range 
and $E_{\rm av} = 1/2(E_{\rm min} +E_{\rm max})$.
For simplicity we have assumed that the systematic errors for neutrino mode and for antineutrino mode are the same. 
For our simulation we have fixed the tilt error to a constant value which is 10\% for T2HK/T2HKK and 2.5\% for DUNE corresponding to all the channels. 
We do not vary these tilt pull variables in our analysis \footnote{Note that the values for the tilt variables are chosen such that our simulation results match with the sensitivities as reported
in the collaboration papers. Thus a variation of the pull variables can also affect the results significantly.}. 
On the other hand the signal and background normalization errors for T2HK, T2HKK and DUNE are considered as variables in our analysis. We have taken them 
to be the same for both appearance and disappearance channel. Thus a systematic uncertainty of $x\%$ implies, a signal and a background normalization error of $x\%$ 
for both appearance and disappearance channel. 

Note that our treatment of the systematics is quite simplistic. In the T2HKK report \cite{T2HKK}, they have considered a total eight pull variables and the values are listed in Table V of the report. 
In DUNE report the systematic uncertainty is taken as 2\% normalization error in appearance channel with respect
to the normalization determined from the disappearance signal, assumed to be known with 5\% uncertainty \cite{Acciarri:2015uup}.


\section{Results for standard three flavor case}
\label{sec3}

In this section we will present the CP violation, hierarchy and octant sensitivity of T2HK, T2HKK and DUNE as a function of the systematic errors. In generating these results
the parameters $\theta_{12}$, $\theta_{13}$, $\Delta m^2_{21}$ and $\Delta m^2_{31}$ are kept fixed close to their best values as obtained in global fits 
\cite{Forero:2014bxa,Esteban:2016qun,Capozzi:2013csa} in both the true and test spectrum. To ensure that the wrong hierarchy minima occurs at a correct value, we have used the
effective formula $\Delta m^2_{\mu\mu}$ given by:
\begin{eqnarray} \nonumber
 \Delta m^2_{31} &=& \Delta m^2_{\mu \mu} + (\cos^2\theta_{12} \\ &-& \cos\dcp \sin \theta_{13} \sin2\theta_{12} \tan\theta_{23})\Delta m^2_{21}
\end{eqnarray}
This is because in the three-flavor scenario, the hierarchy degeneracy does not correspond to $P_{\mu\mu}(\Delta m^2_{31}) = P_{\mu\mu}(-\Delta m^2_{31})$ but it occurs for 
$P_{\mu\mu}(\Delta m^2_{\mu\mu}) = P_{\mu\mu}(-\Delta m^2_{\mu\mu})$ \cite{Nunokawa:2005nx,Raut:2012dm}.
For CP violation study we have taken the true $\theta_{23}=45^\circ$ and marginalized over
$\theta_{23}$, $\dcp$ and hierarchy in the test. For hierarchy sensitivity the true value of $\theta_{23}$ is $45^\circ$ and the true value of $\dcp$ is $\pm 90^\circ$. In this case we have marginalized over
$\theta_{23}$ and $\dcp$ in the test. For octant sensitivity the true value of $\theta_{23}$ is assumed to be $42^\circ$ for the lower octant and $48^\circ$ for the higher octant. 
These values of $\theta_{23}$ are the closest to the best-fit values as obtained by the global fits.
The true value of
$\dcp$ is $- 90^\circ$. For octant sensitivity we have marginalized over $\dcp$ and hierarchy in the test.

\subsection{CP Violation Sensitivity}

\begin{figure*}
\hspace{-30pt}
\includegraphics[scale=1.1]{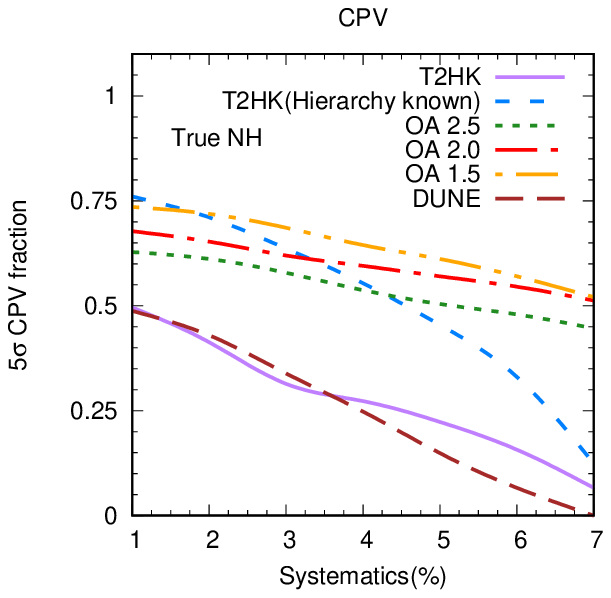}
\hspace{-80pt}
\includegraphics[scale=1.1]{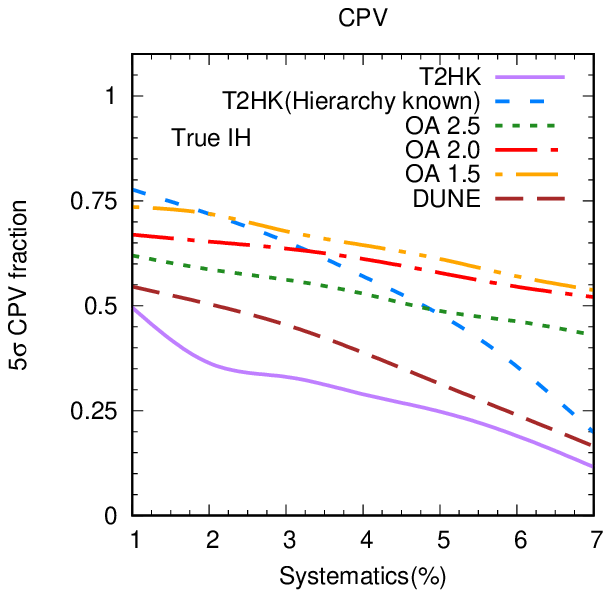}
\caption{Fraction of the true $\dcp$ for which CP violation can be discovered for $5 \sigma$ vs systematics. In all the panels the true $\theta_{23}$ is $45^\circ$. 
Left panel is for NH and right panel is for IH.}
\label{fig1}
\end{figure*}

\begin{figure*}
\hspace{-30pt}
\includegraphics[scale=1.1]{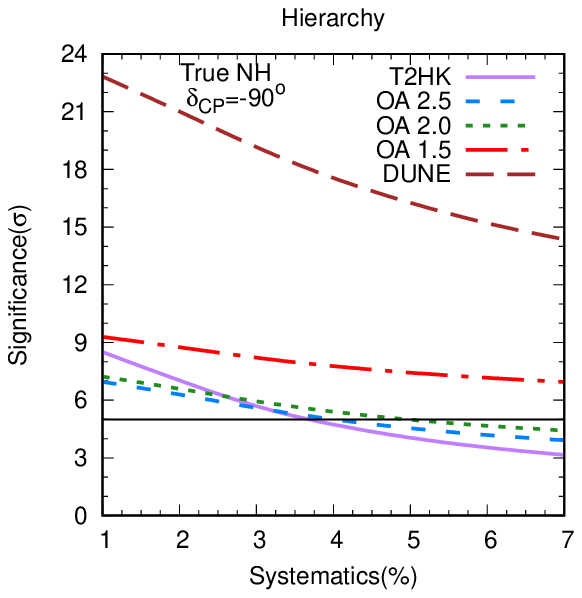}
\hspace{-80pt}
\includegraphics[scale=1.1]{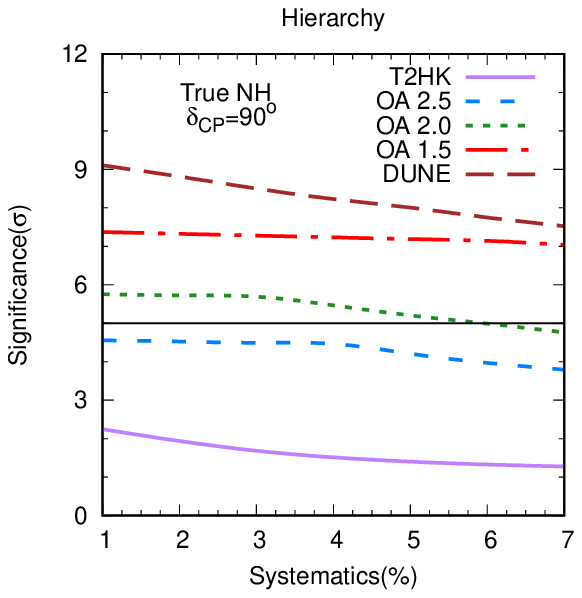} \\
\hspace{-30pt}
\includegraphics[scale=1.1]{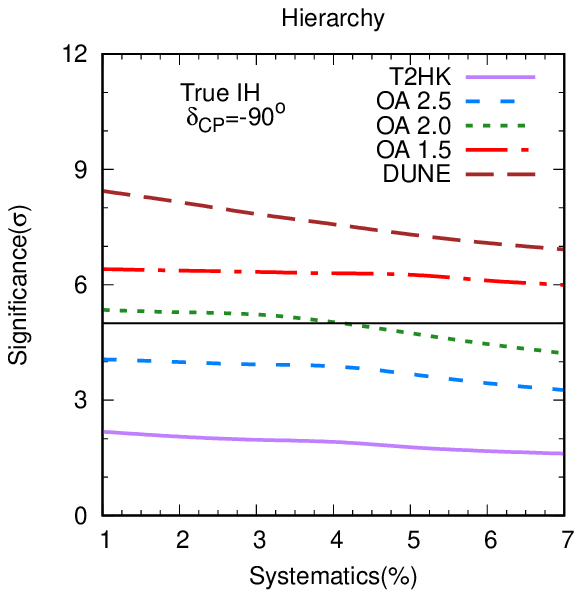}
\hspace{-80pt}
\includegraphics[scale=1.1]{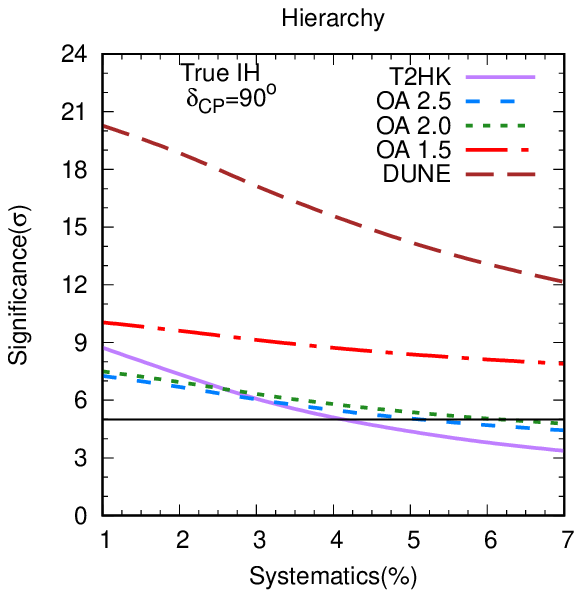}
\caption{Hierarchy sensitivity vs systematics. In all the panels the true value of $\theta_{23}$ is $45^\circ$. The true $\dcp$ is $-90^\circ$ for the left column and $+90^\circ$ for the right column. 
The upper row corresponds to the true NH and the lower row corresponds to the true IH. }
\label{fig2}
\end{figure*}

CP violation (CPV) discovery potential of an experiment is defined by its capability to distinguish a true value of $\dcp$ other than $0^\circ$ or $180^\circ$. In Fig. \ref{fig1}, we have plotted the fraction of the
true $\dcp$ for which CPV can be discovered at $5 \sigma$ vs systematics. The left panel is for normal hierarchy and the right panel is for inverted hierarchy. From the figures we see that
the CPV coverage of DUNE falls from 50\% (55\%) to 0\% (15\%) when the systematics is varied from 1\% to 7\% for NH (IH). For T2HK, the numbers are 50\% (50\%) to 5\% (10\%) if the hierarchy is unknown and 
75\% (80\%) to 15\% (20\%) if hierarchy is known for NH (IH). For $2.5^\circ$
off-axis configuration of T2HKK, the sensitivity falls from 65\% to 45\% for both the hierarchies. For $2.0^\circ$ ($1.5^\circ$) off-axis configurations it falls from 70\% (75\%) to 50\% (50\%) 
in both NH and IH. Thus from these numbers we understand that
the dependence of the sensitivity of T2HK on the systematic errors is stronger than that of T2HKK.
This can be understood in the following way. For T2HK the baseline length is small compared to that of T2HKK. 
Now as the flux drops in proportion to $1/L^2$ (where $L$ is the baseline length) T2HK has much larger event samples at the detector than the T2HKK experiment does and 
thus it is more sensitive to the systematic errors. From the figure
we also see that when hierarchy is unknown then the solid curve (T2HK) always lies below the T2HKK curves. This is because of the presence of degeneracies which restrict the CPV sensitivity of T2HK. However
if the hierarchy is known then we see that the CPV discovery sensitivity of T2HK is better than the $2.5^\circ$ configuration of T2HKK if the systematics is less than 4\% and better than all
the three configurations of T2HKK if the systematics is less than 1\% for both the hierarchies. This is one of the most important findings of our work. From the figures we also note that the CPV discovery
potential of DUNE is comparable to T2HK in NH and slightly higher than T2HK in IH for all the values of systematic errors.

\subsection{Hierarchy Sensitivity}

\begin{figure*}
\begin{tabular}{lr}
\hspace{-30pt}
\includegraphics[width=0.5\textwidth]{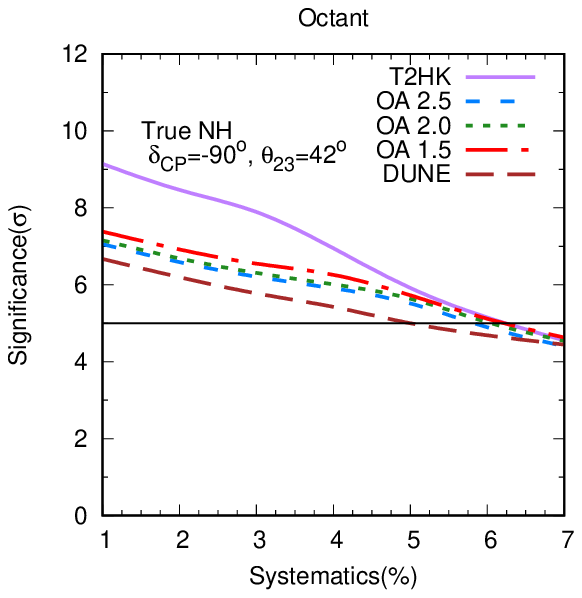}
\hspace{-80pt}
\includegraphics[width=0.5\textwidth]{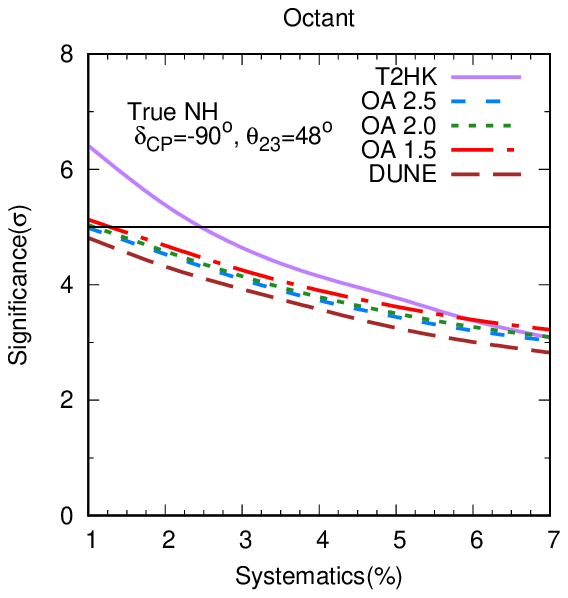} \\
\hspace{-30pt}
\includegraphics[width=0.5\textwidth]{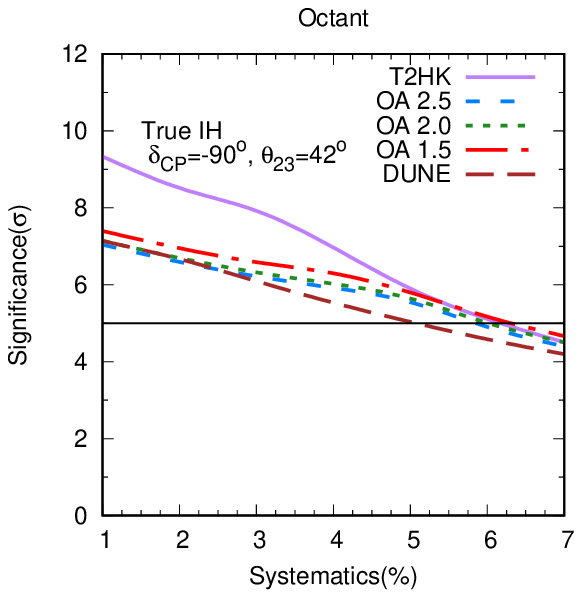}
\hspace{-80pt}
\includegraphics[width=0.5\textwidth]{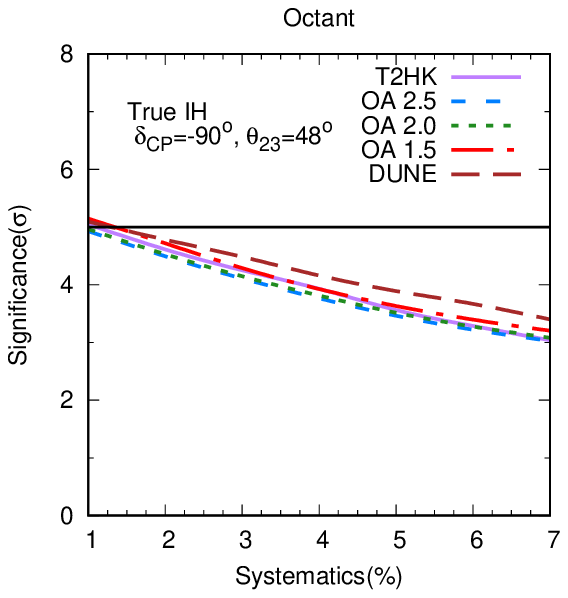}
\end{tabular}
\caption{Octant sensitivity vs systematics. The upper row corresponds to NH and the lower row corresponds to IH. Sensitivities are given for two values of $\theta_{23}$ i.e., $42^\circ$ 
and $48^\circ$ and for $\dcp =-90^\circ$.
}
\label{fig3}
\end{figure*}

The hierarchy sensitivity of an experiment is defined by its capability to rule out the wrong hierarchy solutions. In Fig. \ref{fig2}, we have plotted the hierarchy sensitivity
of T2HK, T2HKK and DUNE vs the 
systematic error. The top row is for normal hierarchy and the bottom row is for inverted hierarchy. In each row the left panel is for $\dcp=-90^\circ$ and the right panel is for $\dcp=90^\circ$.
The horizontal thin solid line corresponds to $5 \sigma$ C.L.
For determination of hierarchy, the favorable combinations of $\dcp$ and hierarchy are \{NH, $-90^\circ$\} and \{IH, $+90^\circ$\}.
These combinations are favorable in determination of the hierarchy because there are no degeneracy for these combinations.
On the other hand, the other two combinations
\{NH, $90^\circ$\} and \{IH, $-90^\circ$\} are not favorable to determine the hierarchy because they
suffers from degeneracy. 

First let us discuss the hierarchy sensitivity for the unfavorable region (top right and bottom left panels). In these case we see that all the curves are almost flat and does not vary much with respect to 
the systematics. 
This is because the sensitivity in the unfavorable parameter space are limited due to the existence of parameter degeneracy and hence they are not dominated by the systematics. From the plots we see that
the sensitivity of DUNE is slightly better than the $1.5^\circ$ off-axis configuration of T2HKK. The sensitivity of DUNE falls from $9 \sigma$ to $7.5 \sigma$ (8.5 $\sigma$ to 7 $\sigma$) in NH (IH) 
as the systematics vary from 1\% to 7\%. For the T2HKK configuration of $1.5^\circ$ corresponding sensitivities are 7.5 $\sigma$ to 7 $\sigma$ (6.5 $\sigma$ to 6 $\sigma$) for NH (IH). 
For the off-axis configurations $2.5^\circ$ of T2HKK,  the sensitivity falls from 4.5 $\sigma$ to $4 \sigma$ (4 $\sigma$ to 3 $\sigma$) in NH (IH). For T2HKK configuration of $2.0^\circ$ the numbers are
6 $\sigma$ to 5 $\sigma$ (5.5 $\sigma$ to 4 $\sigma$) for NH (IH). 
Here we find that the sensitivity of T2HK is the lowest for all the values of the systematics and it remains close to $2 \sigma$ for both the hierarchies.

For the favorable combinations (top left and bottom right panels) we see that the sensitivity of DUNE is maximal among all the setups for all 
the values of the systematics
and the sensitivity falls from $23 \sigma$ (20 $\sigma$) to 14 $\sigma$ (12 $\sigma$) for NH (IH) as the systematics varies from 1\% to 7\%. For T2HK the numbers are 9 $\sigma$ to 3 $\sigma$ for both the 
hierarchies. On the other hand for $1.5^\circ$ ($2.5^\circ$ and $2.0^\circ$) configurations of T2HKK, the sensitivity falls from 9 $\sigma$ to 7 $\sigma$ (7 $\sigma$ to 4 $\sigma$) in NH
and 10 $\sigma$ to 8 $\sigma$ (7.5 $\sigma$ to 4.5 $\sigma$) in IH.
From the plots we also see that the sensitivity of T2HK is better than the $2.0^\circ$ and $2.5^\circ$ configurations of T2HKK if the systematics is less than 2\% and 
for NH the sensitivity of T2HK becomes comparable to $1.5^\circ$ off axis configurations of T2HKK if the systematics is less than 1\%. This is also one of the remarkable findings of our work. 
This also shows that if nature 
choose the true hierarchy to be normal and $\dcp=-90^\circ$ then T2HK is sufficient to determine the hierarchy sensitivity at 5 $\sigma$ confidence level if the systematics is less than 3.5\%.
In these plots we also see that
the curves corresponding to T2HK and DUNE are steeper than those for T2HKK. As explained earlier, because of the large number of event samples at the far detector, the effect of systematics is more for T2HK as
compared to T2HKK although both of their sensitivities lies around $4 \sigma$ to $8 \sigma$. On the other hand, the large variation in the sensitivity of DUNE with respect to the systematic errors are due to its
large matter effect and higher hierarchy sensitivity as compared to T2HK and T2HKK.

In all the panels of Fig. \ref{fig2}, we see that among the three configurations of the T2HKK experiment, the hierarchy sensitivity is best for $1.5^\circ$ off-axis flux and as the detector is moved to higher off-axis
angles, the sensitivity decreases. This is because the $1.5^\circ$ off-axis flux covers more of the first oscillation maxima where the hierarchy sensitivity is maximum \cite{Ballett:2016daj}.

\begin{figure*}
\begin{tabular}{lr}
\vspace{-0.2 in}
\includegraphics[width=0.5\textwidth]{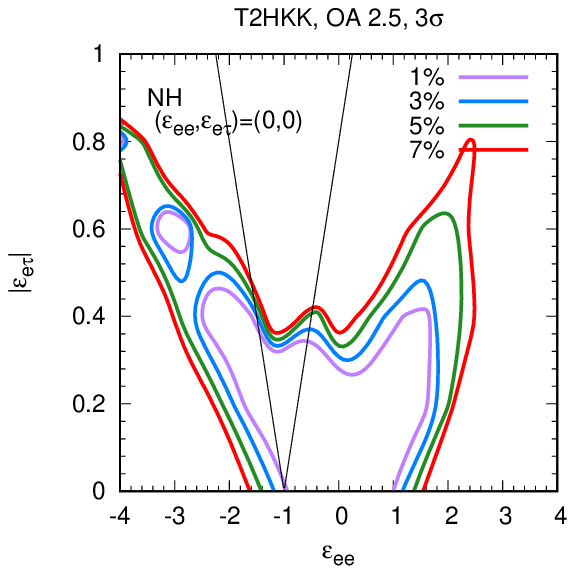}
\hspace*{-1.0in}
\includegraphics[width=0.5\textwidth]{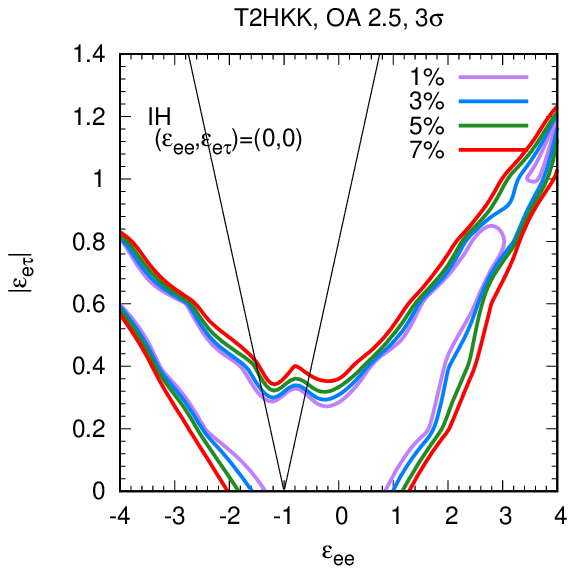} \\
\vspace{-0.2 in}
\includegraphics[width=0.5\textwidth]{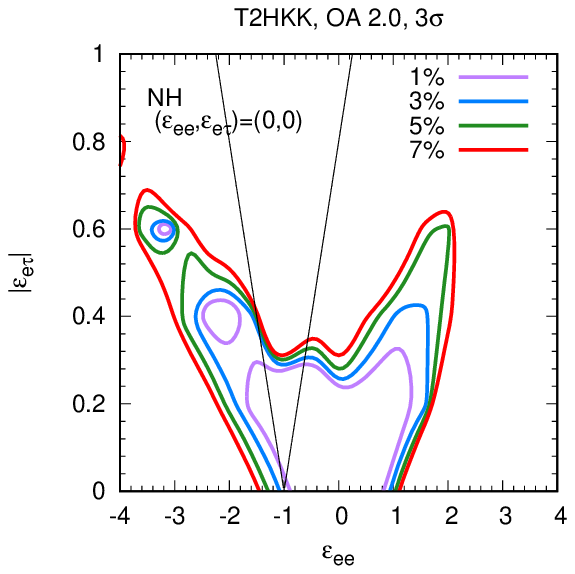}
\hspace*{-1.0in}
\includegraphics[width=0.5\textwidth]{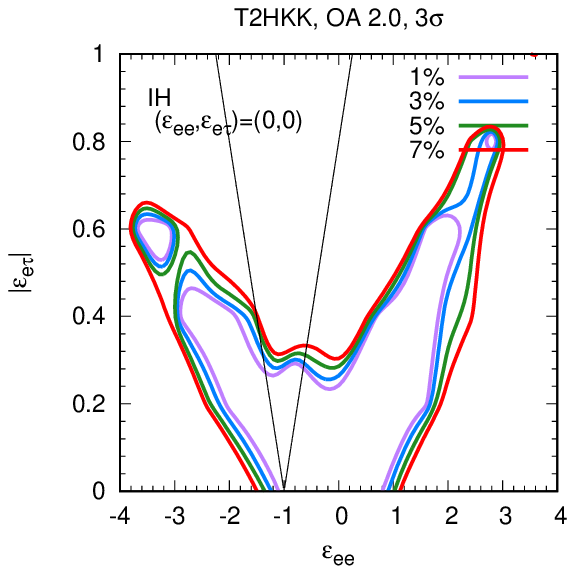} \\ 
\vspace{-0.2 in}
\includegraphics[width=0.5\textwidth]{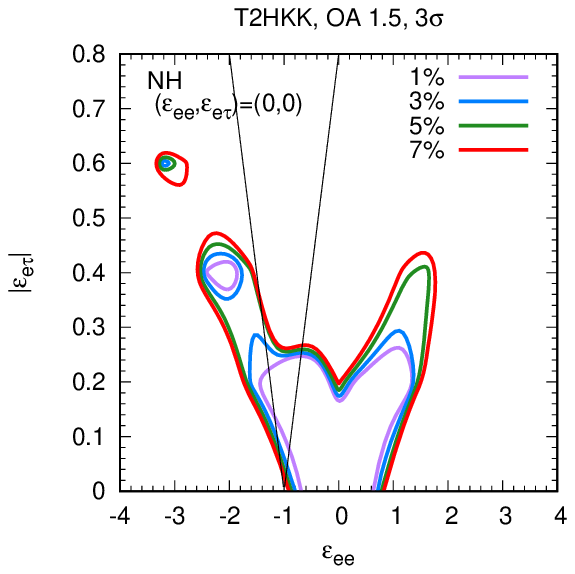}
\hspace*{-1.0in}
\includegraphics[width=0.5\textwidth]{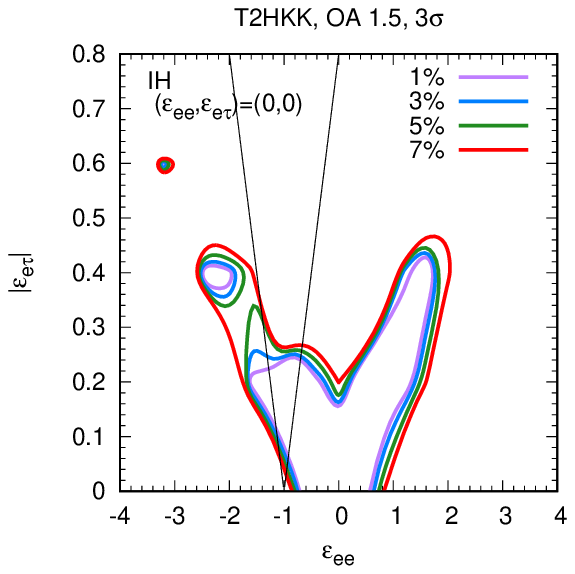} \\ 
\includegraphics[width=0.5\textwidth]{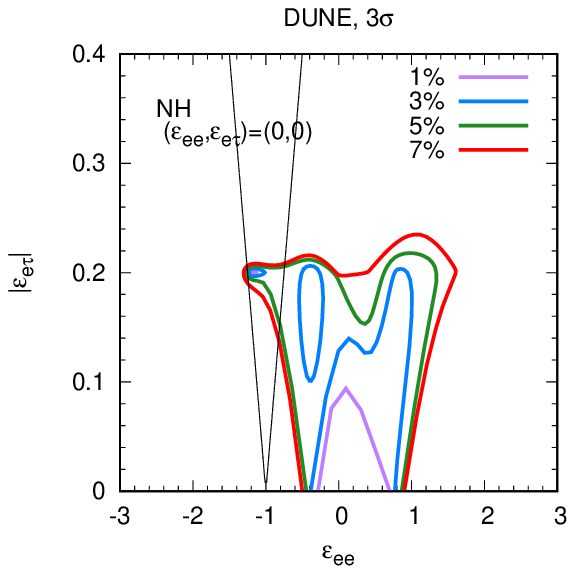}
\hspace*{-1.0in}
\includegraphics[width=0.5\textwidth]{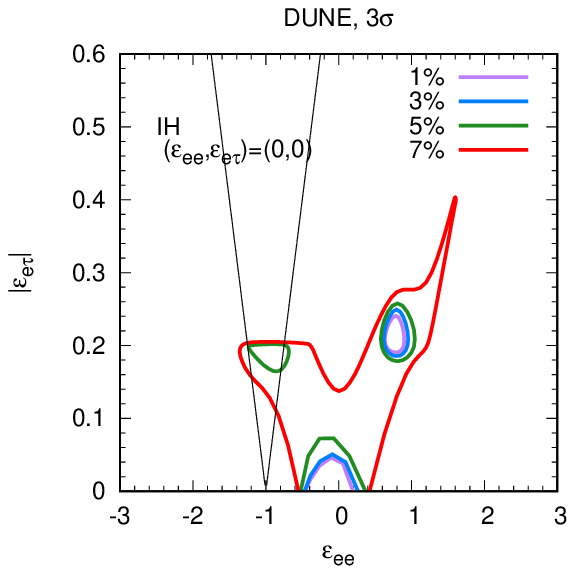} 
\end{tabular}
\caption{Sensitivity for NSI parameters for four values of the systematic errors. Left column is for NH and right column is for IH.
The thin solid diagonal straight line stands for the bound
$|\tan\beta| = |\epsilon_{e\tau}/(1+\epsilon_{ee})|\lesssim
0.8$\,\cite{Fukasawa:2015jaa} at 3$\sigma$ from the current
atmospheric data by Superkamiokande.
First three rows correspond to the $2.5^\circ$, $2.0^\circ$ and $1.5^\circ$
off-axis configurations 
for T2HKK respectively. The fourth row is for DUNE.
For interpretation of the references to color in this figure, the reader is
referred to the web version of this article.}
\label{fig4}
\end{figure*}

\subsection{Octant Sensitivity}

The octant sensitivity of an experiment is defined by its capability to rule out the wrong octant solution. In Fig. \ref{fig3} we have given the octant sensitivity
of T2HK, T2HKK and DUNE as a function
of the systematics. The upper panels are for normal hierarchy and the lower panels are for inverted hierarchy. In each row, the left panel corresponding to LO and the right panel corresponds to HO.
The horizontal thin solid line corresponds to $5 \sigma$ C.L.
All the panels are
for $\dcp=-90^\circ$. Although we have shown our results for only $\dcp=-90^\circ$, we have checked that the conclusion remains the same for all the other values of $\dcp$.

For the lower octant (top left and bottom left panels) we see that if the systematics is less than 4\% than the octant sensitivity of the T2HK experiment is better than all the other experiments.
In this case the sensitivity of T2HK falls from $9 \sigma$ to $4.5 \sigma$ for both the hierarchies as the systematics varies from 1\% to 7\%. 
The sensitivities of all the three configurations of T2HKK are similar and the sensitivity falls from $7 \sigma$ to $4.5 \sigma$ for both the hierarchies. 
Among all the experiments sensitivity of DUNE is the lowest in NH and
the sensitivity varies from $7 \sigma$ to $4.5 \sigma$ for both the hierarchies.
In these plots we see that the dependence of sensitivities on the systematic errors are similar for T2HKK and DUNE whereas the curves for T2HK are steeper than T2HKK and DUNE. This is because unlike the hierarchy
sensitivity, the octant sensitivity of T2HKK and DUNE are similar. On the other hand due to the larger event sample of T2HK, the octant sensitivity of T2HK is greater than T2HKK and DUNE and thus more affected by the
systematic uncertainties.

Now lets discuss the sensitivities for HO (top right and bottom right panels). In these case we see that except T2HK in NH, 
the slope of all the curves are almost equal.
This can be understood in the following way. In general the octant sensitivity in HO is poorer than the LO as the denominator in the $\chi^2$ for HO is higher than LO. Thus it is natural that the effect of systematics
will be less in HO as compared to LO.
In this case the sensitivity of T2HK is the best among all the other 
setups for almost any value of the systematics in NH (thus more affected by systematics) and in IH the sensitivity of T2HK is the same as that of all the three configurations of T2HKK and DUNE
(thus the sensitivity of the all five setups depends similarly on systematics).
Here the sensitivity falls from $6.5 \sigma$ to $3 \sigma$ for NH as the systematics varies from 1\% to 7\%. 
The dependence of the systematics for all the three configurations of T2HKK is
almost the same and they vary from 5\% to 4\% for both the hierarchies. For NH, the sensitivity of DUNE is worse than all the other setups for any value of the systematics
and it varies from $5 \sigma$ to $3 \sigma$.

\begin{figure*}
\begin{tabular}{lr}
\vspace{-0.2 in}
\includegraphics[width=0.5\textwidth]{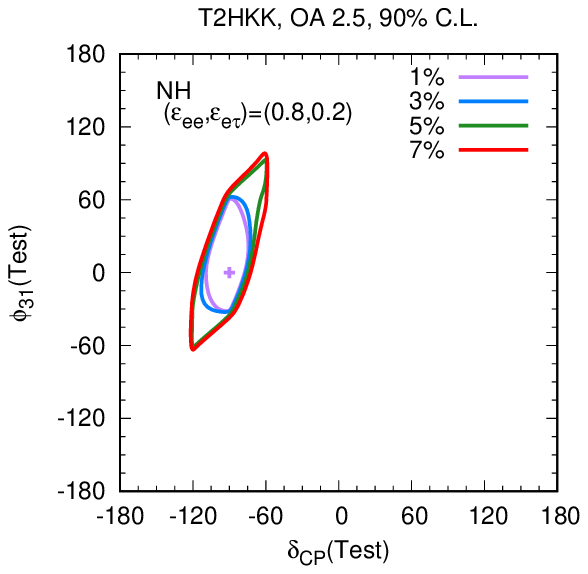}
\hspace*{-1.0in}
\includegraphics[width=0.5\textwidth]{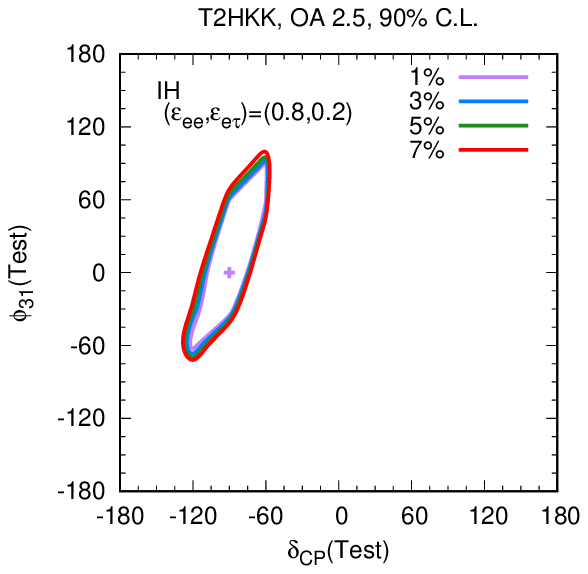} \\
\vspace{-0.2 in}
\includegraphics[width=0.5\textwidth]{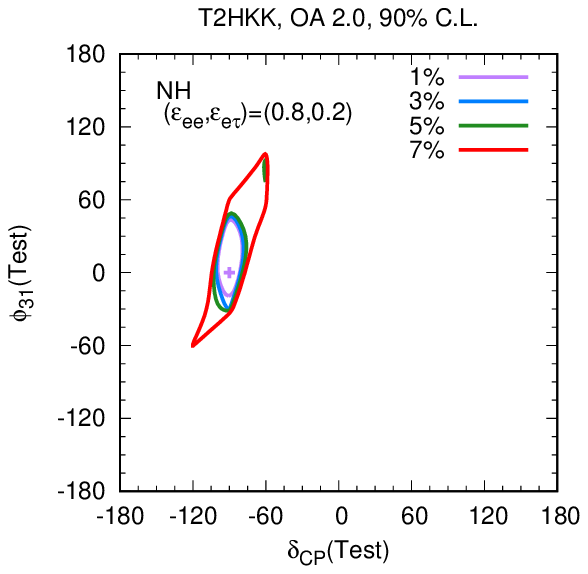}
\hspace*{-1.0in}
\includegraphics[width=0.5\textwidth]{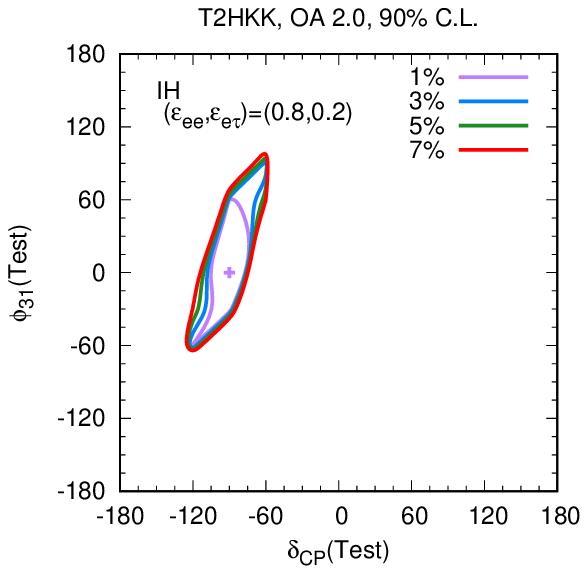} \\ 
\vspace{-0.2 in}
\includegraphics[width=0.5\textwidth]{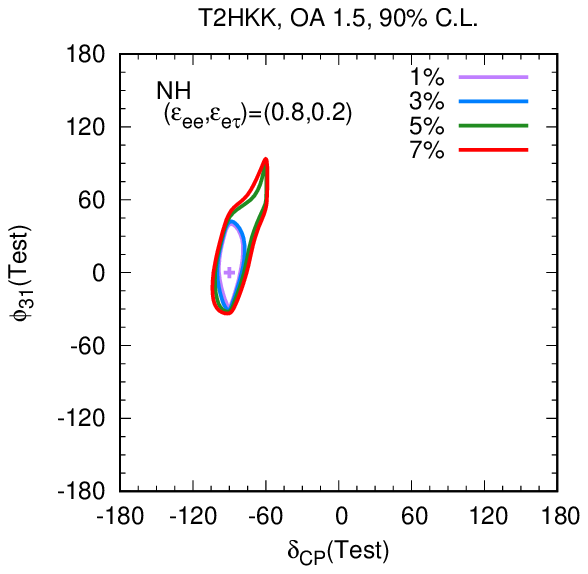}
\hspace*{-1.0in}
\includegraphics[width=0.5\textwidth]{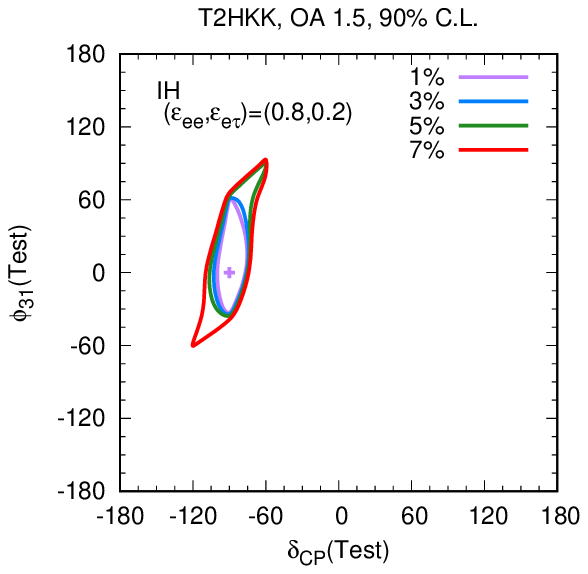} \\ 
\includegraphics[width=0.5\textwidth]{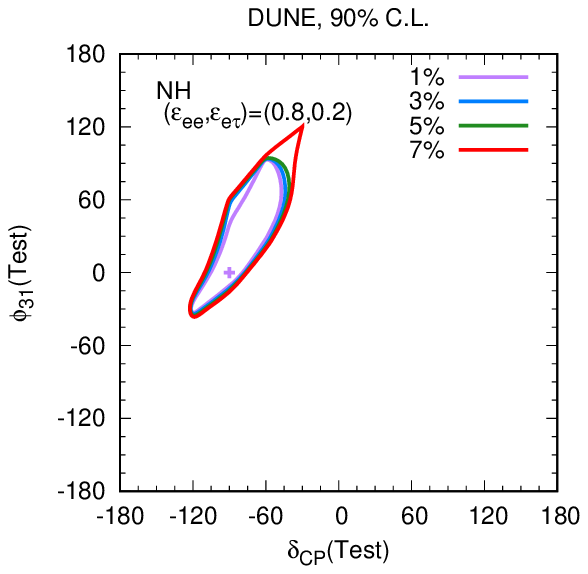}
\hspace*{-1.0in}
\includegraphics[width=0.5\textwidth]{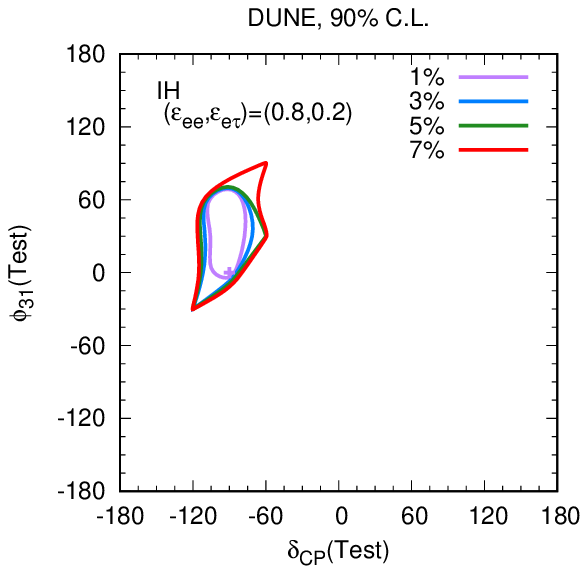} 
\end{tabular}
\caption{Sensitivity to the CP phases for four values of the systematic errors. 
Left column is for NH and right column is for IH. First three rows correspond to the $2.5^\circ$, $2.0^\circ$ and $1.5^\circ$ off-axis configurations 
for T2HKK respectively. The fourth row is for DUNE.
For interpretation of the references to color in this figure, the reader is
referred to the web version of this article.}
\label{fig5}
\end{figure*}

\section{Results for NSI}
\label{sec4}
In this section we will study the effect of the systematics error in constraining the NSI parameters.
The nonstandard interaction in neutrino propagation can arise from the following four-fermion interaction:
\begin{eqnarray}
{\cal L}_{\mbox{\rm\scriptsize eff}}^{\mbox{\tiny{\rm NSI}}} 
=-2\sqrt{2}\, \epsilon_{\alpha\beta}^{ff'P} G_F
\left(\overline{\nu}_{\alpha L} \gamma_\mu \nu_{\beta L}\right)\,
\left(\overline{f}_P \gamma^\mu f_P'\right),
\label{NSIop}
\end{eqnarray}
where $f_P$ and $f_P'$ correspond to fermions with chirality $P$,
$\epsilon_{\alpha\beta}^{ff'P}$ is a dimensionless constant and $G_F$ is the Fermi coupling constant. In the presence of NSI the MSW matter potential takes the following form:
\begin{eqnarray}
{\cal A} \equiv
\sqrt{2} G_F N_e \left(
\begin{array}{ccc}
1+ \epsilon_{ee} & \epsilon_{e\mu} & \epsilon_{e\tau}\\
\epsilon_{\mu e} & \epsilon_{\mu\mu} & \epsilon_{\mu\tau}\\
\epsilon_{\tau e} & \epsilon_{\tau\mu} & \epsilon_{\tau\tau}
\end{array}
\right),
\label{matter-np}
\end{eqnarray}
where $\epsilon_{\alpha\beta}$ is defined by
\begin{equation}
\epsilon_{\alpha\beta}\equiv\sum_{f=e,u,d}\frac{N_f}{N_e}\epsilon_{\alpha\beta}^{f}\,.
\end{equation}
$N_f~(f=e, u, d)$ is the number densities of fermions $f$.
Here we defined the NSI parameters as
$\epsilon_{\alpha\beta}^{fP}\equiv\epsilon_{\alpha\beta}^{ffP}$ and
$\epsilon_{\alpha\beta}^{f}\equiv\epsilon_{\alpha\beta}^{fL}+\epsilon_{\alpha\beta}^{fR}$. The present $90\%$ bounds of the NSI parameters are given by\,\cite{Davidson:2003ha,Biggio:2009nt} 
\begin{eqnarray}
\hspace{-25pt}
&{\ }&
\left(
\begin{array}{lll}
|\epsilon_{ee}| < 4 \times 10^0 & |\epsilon_{e\mu}| < 3\times 10^{-1}
& |\epsilon_{e\tau}| < 3 \times 10^0\\
&  |\epsilon_{\mu\mu}| < 7\times 10^{-2}
& |\epsilon_{\mu\tau}| < 3\times 10^{-1}\\
& & |\epsilon_{\tau\tau}| < 2\times 10^1
\end{array}
\right).
\label{epsilon-m}
\end{eqnarray}
Thus we understand that the bounds on $\epsilon_{\alpha \mu}$ where $\alpha = e$, $\mu$, $\tau$ are stronger than the $\epsilon_{ee}$, $\epsilon_{e\tau}$ and $\epsilon_{\tau\tau}$. 
One additional bound comes from the high-energy atmospheric data which relates the parameters $\epsilon_{\tau\tau}$ and $\epsilon_{e \tau}$ as \cite{Friedland:2004ah,Friedland:2005vy}
\begin{eqnarray}
\epsilon_{\tau\tau} \simeq \frac{|\epsilon_{e\tau}|^2}{1+\epsilon_{ee}}\,.
\label{eq:ansatz_a}
\end{eqnarray}
In Ref.\,\cite{Oki:2010uc} it was shown that,
in the high-energy behavior of the
disappearance oscillation probability
\begin{eqnarray}
&{\ }&\hspace{-20pt}
1-P(\nu_\mu\rightarrow\nu_\mu)
\nonumber\\
&{\ }&\hspace{-30pt}
\simeq
c_0 + c_1\frac{\Delta m^2_{31}/2E}{\sqrt{2} G_F N_e} 
+ {\cal O}\left[\left(\frac{\Delta m^2_{31}/2E}{\sqrt{2} G_F N_e}\right)^2\right],
\label{expansion}
\end{eqnarray}
in the presence of the matter potential (\ref{matter-np}),
$|c_0| \ll 1$ and $|c_1| \ll 1$ imply
$|\epsilon_{e\mu}| \ll 1$,
$|\epsilon_{\mu\mu}| \ll 1$,
$|\epsilon_{\mu\tau}| \ll 1$
and
$|\epsilon_{\tau\tau} -
|\epsilon_{e\tau}|^2/ \left( 1 + \epsilon_{ee} \right)|\ll 1$.
In deriving Eq.\,(\ref{expansion}), it is assumed that
$|1 + \epsilon_{ee}|$ is not very small.
$1 + \epsilon_{ee} = 0$ is the abnormal region
where the $e-e$ component of the matter effect vanishes \cite{Deepthi:2016erc}, and 
the approximation (\ref{eq:ansatz_a}) becomes
invalid only in the
neighborhood of $1 + \epsilon_{ee} = 0$.
As is explained in Appendix A in Ref.\,\cite{Fukasawa:2016nwn}, 
the region $1 + \epsilon_{ee} = 0$
corresponds to the one
$\epsilon_D = 1/6$
in the parametrization of the solar neutrino 
analysis\,\cite{Gonzalez-Garcia:2013usa}.
The result in Ref.\,\cite{Gonzalez-Garcia:2013usa}
shows that the region near $\epsilon_D = 1/6$
is excluded by the solar neutrino and KamLAND data
at more than 3$\sigma$.
The region in which the ansatz (\ref{eq:ansatz_a})
may not be a good approximation is therefore
excluded by the solar neutrino and KamLAND data,
so Eq. (\ref{eq:ansatz_a}) can be justified
by implicitly assuming the prior from the analysis of
the solar neutrino and KamLAND data.\footnote{
In Ref.\,\cite{Fukasawa:2015jaa} is was concluded that
the best-fit point in the $(\epsilon_{ee}, |\epsilon_{e\tau}|)$
plane is given by (-1.0,0.0) using the Superkamiokande
atmospheric neutrino data for 4438 days.  However,
the allowed region within 2$\sigma$
is quite wide in the result of Ref.\,\cite{Fukasawa:2015jaa}
because the analysis is based on the energy rate only,
as the energy spectrum information is not available.
Hence the the best-fit value
$(\epsilon_{ee}, |\epsilon_{e\tau}|)$ = (-1.0,0.0) is only a qualitative estimate.}
Furthermore, from the Super-Kamiokande data\,\cite{Fukasawa:2015jaa}, we get
\begin{eqnarray}
\left|\frac{\epsilon_{e\tau}}{1+\epsilon_{ee}}\right|
\lesssim 0.8\quad\mbox{\rm at}~3\sigma\,.
\label{tanb}
\end{eqnarray}
Keeping these facts in mind, we perform our analysis
with the following ansatz:
\begin{eqnarray}
{\cal A} = \sqrt{2} G_F N_e\left(
\begin{array}{ccc}
1+\epsilon_{ee}&0&\epsilon_{e\tau}\cr
0&0&0\cr
\epsilon_{e\tau}^\ast&0&|\epsilon_{e\tau}|^2/(1+\epsilon_{ee})
\end{array}\right)\,.
\label{ansatz}
\end{eqnarray}
Thus the free parameters are $\epsilon_{ee}$, $|\epsilon_{e \tau}|$ and 
arg($\epsilon_{e\tau}) = \phi_{31}$.

Note that in our earlier work, we have studied the sensitivity of the T2HKK experiment to the nonstandard interactions in a similar fashion \cite{Fukasawa:2016lew}. But at that time,
the experimental details for the various T2HKK setups were not available and thus we have used the T2HK setup of \cite{Abe:2014oxa} and scaled those events at 1100 km. But in the present analysis,
we have used the configurations of the T2HKK as outlined in \cite{T2HKK}. In the previous work the ratio of the neutrino and antineutrino run was 1:1 but in the present analysis, we have taken the ratio to be 1:3
following the T2HKK report. But the major difference of the present work as compared to the earlier work lies in the fact that, here we have presented our results for four different sets of systematic errors while 
in the previous work only one set of systematic errors was considered.

\subsection{Constraining the NSI parameters}

In Fig. \ref{fig4}, we have plotted the sensitivity for the NSI parameters for different four values of the systematic errors in the $\epsilon_{ee}$ (test) vs $|\epsilon_{e\tau}|$ (test) plane. The true values of both
the parameters are zero. Thus this plots describes the potential of these future beam-based experiments to put bounds on the NSI parameters. The true value of $\dcp$ is taken as $-90^\circ$ and the true 
value of $\phi_{31}$ is zero. Both the parameters are marginalized over in the test. The true value of $\theta_{23}$ is $45^\circ$ and this parameter is marginalized over in the test. The parameters $\theta_{13}$, 
$\theta_{12}$, $\Delta m^2_{21}$ and $\Delta m^2_{31}$ are kept fixed in both the true and test spectrums. Hierarchy is assumed to be known for these plots. 
The left (right) column of Fig. \ref{fig4} corresponds to NH (IH). 
The first three rows correspond to the $2.5^\circ$, $2.0^\circ$ and $1.5^\circ$ off-axis configurations of T2HKK respectively and the fourth row is for DUNE.
In each panel the back solid curve correspond to the bound $|\epsilon_{e\tau}/(1+\epsilon_{ee})|
\lesssim 0.8~\mbox{\rm at}~3\sigma$. 

From the figure we see that for the setups (OA 2.5, NH) and (OA 1.5, NH), the sensitivities corresponding to the 1\% and 3\% systematic uncertainties
are similar so as the sensitivities corresponding to 5\% and 7\%. An improvement of the systematics from 5\% to 3\% improves the sensitivity significantly. 
For (OA 2.5, IH) the sensitivity for all the four cases of the systematic
 uncertainties are similar. Thus in this case the systematic uncertainties does not play a significant role. 
 For (OA 2.0, NH) and (DUNE, NH) the sensitivity for the NSI parameters gets improved as the systematics errors are lowered from
 7\% to 1\%. For (OA 2.0, IH), (OA 1.5, IH) and (DUNE, IH) the sensitivity corresponding to the 1\%, 3\% and 5\% systematic errors are similar. Among the three setups of T2HKK experiment, the configuration with 
 $1.5^\circ$ off-axis flux, covers maximum area in the probability spectrum in the higher energy region and the configuration with $2.5^\circ$ covers the minimum area in the probability spectrum in the lower energy 
 region. For this reason the capability of constraining the NSI parameters is maximum for $1.5^\circ$ off-axis configuration and minimum for $2.5^\circ$ off-axis configuration. For similar reason, the effect of
 systematics in $1.5^\circ$ off-axis configuration is more as compared to the other two setups of T2HKK.

\subsection{Constraining the CP phases}

In Fig. \ref{fig5}, we have plotted the 90\% C.L contours in the $\dcp$ (test) - $\phi_{31} $ (test) plane for four values of the systematic errors for a true value of $\dcp=-90^\circ$ and $\phi_{31}=0^\circ$. 
For these plots we have considered the true values of 
$\epsilon_{ee} = 0.8$ and $|\epsilon_{e\tau}| = 0.2$ and they are marginalized over in the test. Thus these figures corresponding to capability of these future baseline long-baseline
experiments to constrain the CP phases assuming NSI exists in nature. The parameter $\theta_{23}$ is $45^\circ$ in the true spectrum and marginalized over in the test. As for the earlier case 
$\theta_{13}$, 
$\theta_{12}$, $\Delta m^2_{21}$ and $\Delta m^2_{31}$ are kept fixed in both the true and test spectrums. Hierarchy is assumed to be known in all the panels. The left column is for NH and right column is 
for IH. The first three rows corresponds to the $2.5^\circ$, $2.0^\circ$ and $1.5^\circ$ off-axis configurations of T2HKK respectively. 
The fourth row is for DUNE. In each panels the `+' symbol signifies the true values
of $\dcp$ and $\phi_{31}$. 

From the plots we see the following. For the setups (OA 2.5, NH), (OA 1.5, NH) and (OA 1.5, IH) the sensitivities corresponding to the 1\% and 3\% systematic errors are similar so the
sensitivities of 5\% and 7\%. A reduction of the systematics from 5\% to 3\% causes a significant improvement in the sensitivity. For the setup (OA 2.5, IH), the systematic uncertainties does not play any role for the
sensitivity as the result is similar for all the four values of the systematic errors. 
For (OA 2.0, NH), (DUNE, NH) and (DUNE, IH) the sensitivity is similar for the systematic uncertainties of 1\%, 3\% and 5\%. On the other hand for the setup
(OA 2.0, IH), the sensitivity corresponding to 3\%, 5\% and 7\% are quite similar. As explained earlier, the capability of $1.5^\circ$ off-axis configuration in constraining the CP phases is better among the other 
two setups of T2HKK.

\section{Conclusion}
\label{sec5}

In this paper we have done a comparative study of T2HK, T2HKK and DUNE in the standard three flavor scenario and the case with new physics with respect to the systematic errors. 
In the standard oscillation case we have studied the effect of the systematics in determining the unknowns in the neutrino oscillation sector i.e., neutrino mass hierarchy, octant of the mixing angle $\theta_{23}$ 
and CP violation. As a probe of new physics we have analyzed the role of the systematic uncertainties in constraining the parameters of the nonstandard interactions in the neutrino propagation.
In our analysis we have taken four pull variables which are (i) signal normalization, (ii) background normalization, (iii) signal tilt and (iv) background tilt.
We have fixed the tilt errors to a constant value and presented our results as a function of the normalization errors. 
The normalization errors corresponding to signal and background are considered to be equal to each other in our treatment. 
Note that our method of incorporating the systematic errors are quite simplistic in the sense that instead of considering the pull variables for each source of systematics as in \cite{T2HKK}, we have considered
a single pull variable corresponding to an overall normalization factor. But still the results discussed in this work provide an useful guidance on how
the sensitivity of various experiments may depend on the achieved systematic uncertainties.
We find that the variation of the sensitivity with respect to the systematic errors depends on 
the nature of the true parameter space. For a given set of the true parameters, if the sensitivity of a particular experiment is very high, then its sensitivity varies much more with respect to their systematic errors.
Our results also show that the variation of the sensitivity of the T2HK experiment is larger compared to that for the T2HKK setup in the standard oscillation scenario. The major findings of our work are:
(i) If hierarchy is known, then T2HK can have the best CPV discovery sensitivity among all the setups considered in this analysis if the systematics are reduced to within 1\%. 
The CPV discovery potential of T2HK is poor because of the lack of information of mass hierarchy/matter effect. On the other hand, placing a detector at a longer distance may affect the CP sensitivity 
because of the reduction in statistics. Thus, if the information of mass hierarchy comes from a different experiment like DUNE, then our results establish the fact that placing both the detectors at 295 km 
can give the better CP sensitivity than keeping one detector at 295 km and the other at 1100 km, if the systematics are reduced to 1\%.
(ii) For the favorable combinations of the true hierarchy and the true $\dcp$, the hierarchy sensitivity of T2HK will be comparable to the $1.5^\circ$ off-axis configuration of T2HKK 
if the systematics is 1\% and the true hierarchy is normal. But among all 
the setups, the best hierarchy sensitivity will come for DUNE for any value of systematic errors for this favorable parameter space. 
For the unfavorable parameter space, the hierarchy sensitivities are almost constant when the systematic
error is varied but still the best sensitivity comes from the DUNE experiment. 
(iii) Regarding the octant sensitivity, we find that if nature chooses the lower octant, then T2HK will give the best octant sensitivity among all 
the setups if systematic error is less than 4\%
and in the case of the true higher octant, the sensitivity of T2HK is always the best for all the values of the systematics for NH.
On the other hand, in the study of the dependence of the sensitivity to the NSI parameters on systematic errors, we find that different setups respond differently for various values of systematic errors. Our major findings are: 
(i) Apart from the $2.0^\circ$ off-axis configuration in IH i.e., (OA 2.0, IH), capabilities of all the setups are sensitive to systematic errors. 
(ii) The capability of (OA 2.0, NH) and (DUNE, NH) in constraining $\epsilon_{ee}$ and $|\epsilon_{e\tau}|$ gets chronologically improved as the systematics is reduced from 7\% to 1\%.
(iii) For the other setups, the sensitivities corresponding to the 1\% and 3\% systematics errors are similar in constraining $\epsilon_{ee}$ and $|\epsilon_{e\tau}|$.
(iv) In constraining the CP phases, except (OA 2.0, IH), lowering systematic uncertainty below
3\% would not result in a significant improvement in the achieved sensitivity.
Thus from the above discussion, we understand that measurement of the systematics plays an important role in the sensitivity reach of the future high statistics long-baseline experiments. 
Depending on the value of the systematics, the sensitivity of one 
experiment can be better than others. 
Thus the results shown in this work can serve as a guidance for much more realistic studies by experimental groups in planning the future generation long-baseline experiments.


\section*{Acknowledgements}

MG would like to thank Srubabati Goswami for useful discussions. MG also thanks Kaustav Chakraborty for useful discussions on GLoBES. This research was partly supported by a Grant-in-Aid for Scientific
Research of the Ministry of Education, Science and Culture, under
Grants No. 25105009, No. 15K05058, No. 25105001 and No. 15K21734.


\bibliography{t2hkk_sys}

\begin{thebibliography}{72}%
\makeatletter
\providecommand \@ifxundefined [1]{%
 \@ifx{#1\undefined}
}%
\providecommand \@ifnum [1]{%
 \ifnum #1\expandafter \@firstoftwo
 \else \expandafter \@secondoftwo
 \fi
}%
\providecommand \@ifx [1]{%
 \ifx #1\expandafter \@firstoftwo
 \else \expandafter \@secondoftwo
 \fi
}%
\providecommand \natexlab [1]{#1}%
\providecommand \enquote  [1]{``#1''}%
\providecommand \bibnamefont  [1]{#1}%
\providecommand \bibfnamefont [1]{#1}%
\providecommand \citenamefont [1]{#1}%
\providecommand \href@noop [0]{\@secondoftwo}%
\providecommand \href [0]{\begingroup \@sanitize@url \@href}%
\providecommand \@href[1]{\@@startlink{#1}\@@href}%
\providecommand \@@href[1]{\endgroup#1\@@endlink}%
\providecommand \@sanitize@url [0]{\catcode `\\12\catcode `\$12\catcode
  `\&12\catcode `\#12\catcode `\^12\catcode `\_12\catcode `\%12\relax}%
\providecommand \@@startlink[1]{}%
\providecommand \@@endlink[0]{}%
\providecommand \url  [0]{\begingroup\@sanitize@url \@url }%
\providecommand \@url [1]{\endgroup\@href {#1}{\urlprefix }}%
\providecommand \urlprefix  [0]{URL }%
\providecommand \Eprint [0]{\href }%
\providecommand \doibase [0]{http://dx.doi.org/}%
\providecommand \selectlanguage [0]{\@gobble}%
\providecommand \bibinfo  [0]{\@secondoftwo}%
\providecommand \bibfield  [0]{\@secondoftwo}%
\providecommand \translation [1]{[#1]}%
\providecommand \BibitemOpen [0]{}%
\providecommand \bibitemStop [0]{}%
\providecommand \bibitemNoStop [0]{.\EOS\space}%
\providecommand \EOS [0]{\spacefactor3000\relax}%
\providecommand \BibitemShut  [1]{\csname bibitem#1\endcsname}%
\let\auto@bib@innerbib\@empty
\bibitem [{\citenamefont {Magaletti}(2016)}]{t2k}%
  \BibitemOpen
  \bibfield  {author} {\bibinfo {author} {\bibfnamefont {L.}~\bibnamefont
  {Magaletti}},\ }\href@noop {} {\  (\bibinfo {year} {2016})},\ \bibinfo {note}
  {{talk at NOW2016, Otranto, Italy, 4 -- 11 September, 2016.}}\BibitemShut
  {Stop}%
\bibitem [{\citenamefont {Vahle}(2016)}]{nova}%
  \BibitemOpen
  \bibfield  {author} {\bibinfo {author} {\bibfnamefont {P.}~\bibnamefont
  {Vahle}},\ }\href@noop {} {\  (\bibinfo {year} {2016})},\ \bibinfo {note}
  {{talk at Neutrino 2016, 4-9 July, London.}}\BibitemShut {Stop}%
\bibitem [{\citenamefont {Adamson}\ \emph {et~al.}(2017)\citenamefont {Adamson}
  \emph {et~al.}}]{Adamson:2017qqn}%
  \BibitemOpen
  \bibfield  {author} {\bibinfo {author} {\bibfnamefont {P.}~\bibnamefont
  {Adamson}} \emph {et~al.} (\bibinfo {collaboration} {NOvA}),\ }\href@noop {}
  {\bibfield  {journal} {\bibinfo  {journal} {Submitted to: Phys. Rev. Lett.}\
  } (\bibinfo {year} {2017})},\ \Eprint {http://arxiv.org/abs/1701.05891}
  {arXiv:1701.05891 [hep-ex]} \BibitemShut {NoStop}%
\bibitem [{\citenamefont {Prakash}\ \emph {et~al.}(2012)\citenamefont
  {Prakash}, \citenamefont {Raut},\ and\ \citenamefont
  {Sankar}}]{Prakash:2012az}%
  \BibitemOpen
  \bibfield  {author} {\bibinfo {author} {\bibfnamefont {S.}~\bibnamefont
  {Prakash}}, \bibinfo {author} {\bibfnamefont {S.~K.}\ \bibnamefont {Raut}}, \
  and\ \bibinfo {author} {\bibfnamefont {S.~U.}\ \bibnamefont {Sankar}},\
  }\href {\doibase 10.1103/PhysRevD.86.033012} {\bibfield  {journal} {\bibinfo
  {journal} {Phys. Rev.}\ }\textbf {\bibinfo {volume} {D86}},\ \bibinfo {pages}
  {033012} (\bibinfo {year} {2012})},\ \Eprint {http://arxiv.org/abs/1201.6485}
  {arXiv:1201.6485 [hep-ph]} \BibitemShut {NoStop}%
\bibitem [{\citenamefont {Agarwalla}\ \emph {et~al.}(2013)\citenamefont
  {Agarwalla}, \citenamefont {Prakash},\ and\ \citenamefont
  {Sankar}}]{Agarwalla:2013ju}%
  \BibitemOpen
  \bibfield  {author} {\bibinfo {author} {\bibfnamefont {S.~K.}\ \bibnamefont
  {Agarwalla}}, \bibinfo {author} {\bibfnamefont {S.}~\bibnamefont {Prakash}},
  \ and\ \bibinfo {author} {\bibfnamefont {S.~U.}\ \bibnamefont {Sankar}},\
  }\href {\doibase 10.1007/JHEP07(2013)131} {\bibfield  {journal} {\bibinfo
  {journal} {JHEP}\ }\textbf {\bibinfo {volume} {07}},\ \bibinfo {pages} {131}
  (\bibinfo {year} {2013})},\ \Eprint {http://arxiv.org/abs/1301.2574}
  {arXiv:1301.2574 [hep-ph]} \BibitemShut {NoStop}%
\bibitem [{\citenamefont {Barger}\ \emph {et~al.}(2002)\citenamefont {Barger},
  \citenamefont {Marfatia},\ and\ \citenamefont {Whisnant}}]{Barger:2001yr}%
  \BibitemOpen
  \bibfield  {author} {\bibinfo {author} {\bibfnamefont {V.}~\bibnamefont
  {Barger}}, \bibinfo {author} {\bibfnamefont {D.}~\bibnamefont {Marfatia}}, \
  and\ \bibinfo {author} {\bibfnamefont {K.}~\bibnamefont {Whisnant}},\ }\href
  {\doibase 10.1103/PhysRevD.65.073023} {\bibfield  {journal} {\bibinfo
  {journal} {Phys. Rev.}\ }\textbf {\bibinfo {volume} {D65}},\ \bibinfo {pages}
  {073023} (\bibinfo {year} {2002})},\ \Eprint
  {http://arxiv.org/abs/hep-ph/0112119} {arXiv:hep-ph/0112119 [hep-ph]}
  \BibitemShut {NoStop}%
\bibitem [{\citenamefont {Ghosh}\ \emph {et~al.}(2016)\citenamefont {Ghosh},
  \citenamefont {Ghoshal}, \citenamefont {Goswami}, \citenamefont {Nath},\ and\
  \citenamefont {Raut}}]{Ghosh:2015ena}%
  \BibitemOpen
  \bibfield  {author} {\bibinfo {author} {\bibfnamefont {M.}~\bibnamefont
  {Ghosh}}, \bibinfo {author} {\bibfnamefont {P.}~\bibnamefont {Ghoshal}},
  \bibinfo {author} {\bibfnamefont {S.}~\bibnamefont {Goswami}}, \bibinfo
  {author} {\bibfnamefont {N.}~\bibnamefont {Nath}}, \ and\ \bibinfo {author}
  {\bibfnamefont {S.~K.}\ \bibnamefont {Raut}},\ }\href {\doibase
  10.1103/PhysRevD.93.013013} {\bibfield  {journal} {\bibinfo  {journal} {Phys.
  Rev.}\ }\textbf {\bibinfo {volume} {D93}},\ \bibinfo {pages} {013013}
  (\bibinfo {year} {2016})},\ \Eprint {http://arxiv.org/abs/1504.06283}
  {arXiv:1504.06283 [hep-ph]} \BibitemShut {NoStop}%
\bibitem [{\citenamefont {Ghosh}\ \emph {et~al.}(2013)\citenamefont {Ghosh},
  \citenamefont {Thakore},\ and\ \citenamefont {Choubey}}]{Ghosh:2012px}%
  \BibitemOpen
  \bibfield  {author} {\bibinfo {author} {\bibfnamefont {A.}~\bibnamefont
  {Ghosh}}, \bibinfo {author} {\bibfnamefont {T.}~\bibnamefont {Thakore}}, \
  and\ \bibinfo {author} {\bibfnamefont {S.}~\bibnamefont {Choubey}},\ }\href
  {\doibase 10.1007/JHEP04(2013)009} {\bibfield  {journal} {\bibinfo  {journal}
  {JHEP}\ }\textbf {\bibinfo {volume} {04}},\ \bibinfo {pages} {009} (\bibinfo
  {year} {2013})},\ \Eprint {http://arxiv.org/abs/1212.1305} {arXiv:1212.1305
  [hep-ph]} \BibitemShut {NoStop}%
\bibitem [{\citenamefont {Ghosh}\ \emph
  {et~al.}(2014{\natexlab{a}})\citenamefont {Ghosh}, \citenamefont {Ghoshal},
  \citenamefont {Goswami},\ and\ \citenamefont {Raut}}]{Ghosh:2013zna}%
  \BibitemOpen
  \bibfield  {author} {\bibinfo {author} {\bibfnamefont {M.}~\bibnamefont
  {Ghosh}}, \bibinfo {author} {\bibfnamefont {P.}~\bibnamefont {Ghoshal}},
  \bibinfo {author} {\bibfnamefont {S.}~\bibnamefont {Goswami}}, \ and\
  \bibinfo {author} {\bibfnamefont {S.~K.}\ \bibnamefont {Raut}},\ }\href
  {\doibase 10.1103/PhysRevD.89.011301} {\bibfield  {journal} {\bibinfo
  {journal} {Phys. Rev.}\ }\textbf {\bibinfo {volume} {D89}},\ \bibinfo {pages}
  {011301} (\bibinfo {year} {2014}{\natexlab{a}})},\ \Eprint
  {http://arxiv.org/abs/1306.2500} {arXiv:1306.2500 [hep-ph]} \BibitemShut
  {NoStop}%
\bibitem [{\citenamefont {Prakash}\ \emph {et~al.}(2014)\citenamefont
  {Prakash}, \citenamefont {Rahaman},\ and\ \citenamefont
  {Sankar}}]{Prakash:2013dua}%
  \BibitemOpen
  \bibfield  {author} {\bibinfo {author} {\bibfnamefont {S.}~\bibnamefont
  {Prakash}}, \bibinfo {author} {\bibfnamefont {U.}~\bibnamefont {Rahaman}}, \
  and\ \bibinfo {author} {\bibfnamefont {S.~U.}\ \bibnamefont {Sankar}},\
  }\href {\doibase 10.1007/JHEP07(2014)070} {\bibfield  {journal} {\bibinfo
  {journal} {JHEP}\ }\textbf {\bibinfo {volume} {07}},\ \bibinfo {pages} {070}
  (\bibinfo {year} {2014})},\ \Eprint {http://arxiv.org/abs/1306.4125}
  {arXiv:1306.4125 [hep-ph]} \BibitemShut {NoStop}%
\bibitem [{\citenamefont {Ghosh}\ \emph
  {et~al.}(2014{\natexlab{b}})\citenamefont {Ghosh}, \citenamefont {Ghoshal},
  \citenamefont {Goswami},\ and\ \citenamefont {Raut}}]{Ghosh:2014dba}%
  \BibitemOpen
  \bibfield  {author} {\bibinfo {author} {\bibfnamefont {M.}~\bibnamefont
  {Ghosh}}, \bibinfo {author} {\bibfnamefont {P.}~\bibnamefont {Ghoshal}},
  \bibinfo {author} {\bibfnamefont {S.}~\bibnamefont {Goswami}}, \ and\
  \bibinfo {author} {\bibfnamefont {S.~K.}\ \bibnamefont {Raut}},\ }\href
  {\doibase 10.1016/j.nuclphysb.2014.04.013} {\bibfield  {journal} {\bibinfo
  {journal} {Nucl. Phys.}\ }\textbf {\bibinfo {volume} {B884}},\ \bibinfo
  {pages} {274} (\bibinfo {year} {2014}{\natexlab{b}})},\ \Eprint
  {http://arxiv.org/abs/1401.7243} {arXiv:1401.7243 [hep-ph]} \BibitemShut
  {NoStop}%
\bibitem [{\citenamefont {Chatterjee}\ \emph {et~al.}(2013)\citenamefont
  {Chatterjee}, \citenamefont {Ghoshal}, \citenamefont {Goswami},\ and\
  \citenamefont {Raut}}]{Chatterjee:2013qus}%
  \BibitemOpen
  \bibfield  {author} {\bibinfo {author} {\bibfnamefont {A.}~\bibnamefont
  {Chatterjee}}, \bibinfo {author} {\bibfnamefont {P.}~\bibnamefont {Ghoshal}},
  \bibinfo {author} {\bibfnamefont {S.}~\bibnamefont {Goswami}}, \ and\
  \bibinfo {author} {\bibfnamefont {S.~K.}\ \bibnamefont {Raut}},\ }\href
  {\doibase 10.1007/JHEP06(2013)010} {\bibfield  {journal} {\bibinfo  {journal}
  {JHEP}\ }\textbf {\bibinfo {volume} {06}},\ \bibinfo {pages} {010} (\bibinfo
  {year} {2013})},\ \Eprint {http://arxiv.org/abs/1302.1370} {arXiv:1302.1370
  [hep-ph]} \BibitemShut {NoStop}%
\bibitem [{\citenamefont {Ghosh}(2016)}]{Ghosh:2015tan}%
  \BibitemOpen
  \bibfield  {author} {\bibinfo {author} {\bibfnamefont {M.}~\bibnamefont
  {Ghosh}},\ }\href {\doibase 10.1103/PhysRevD.93.073003} {\bibfield  {journal}
  {\bibinfo  {journal} {Phys. Rev.}\ }\textbf {\bibinfo {volume} {D93}},\
  \bibinfo {pages} {073003} (\bibinfo {year} {2016})},\ \Eprint
  {http://arxiv.org/abs/1512.02226} {arXiv:1512.02226 [hep-ph]} \BibitemShut
  {NoStop}%
\bibitem [{\citenamefont {Bharti}\ \emph {et~al.}(2016)\citenamefont {Bharti},
  \citenamefont {Prakash}, \citenamefont {Rahaman},\ and\ \citenamefont
  {Sankar}}]{Bharti:2016hfb}%
  \BibitemOpen
  \bibfield  {author} {\bibinfo {author} {\bibfnamefont {S.}~\bibnamefont
  {Bharti}}, \bibinfo {author} {\bibfnamefont {S.}~\bibnamefont {Prakash}},
  \bibinfo {author} {\bibfnamefont {U.}~\bibnamefont {Rahaman}}, \ and\
  \bibinfo {author} {\bibfnamefont {S.~U.}\ \bibnamefont {Sankar}},\
  }\href@noop {} {\  (\bibinfo {year} {2016})},\ \Eprint
  {http://arxiv.org/abs/1602.03513} {arXiv:1602.03513 [hep-ph]} \BibitemShut
  {NoStop}%
\bibitem [{\citenamefont {Soumya}\ and\ \citenamefont
  {Mohanta}(2016{\natexlab{a}})}]{Soumya:2016aif}%
  \BibitemOpen
  \bibfield  {author} {\bibinfo {author} {\bibfnamefont {C.}~\bibnamefont
  {Soumya}}\ and\ \bibinfo {author} {\bibfnamefont {R.}~\bibnamefont
  {Mohanta}},\ }\href {\doibase 10.1140/epjc/s10052-016-4125-6} {\bibfield
  {journal} {\bibinfo  {journal} {Eur. Phys. J.}\ }\textbf {\bibinfo {volume}
  {C76}},\ \bibinfo {pages} {302} (\bibinfo {year} {2016}{\natexlab{a}})},\
  \Eprint {http://arxiv.org/abs/1605.00523} {arXiv:1605.00523 [hep-ph]}
  \BibitemShut {NoStop}%
\bibitem [{\citenamefont {Abe}\ \emph {et~al.}(2014)\citenamefont {Abe} \emph
  {et~al.}}]{Abe:2014oxa}%
  \BibitemOpen
  \bibfield  {author} {\bibinfo {author} {\bibfnamefont {K.}~\bibnamefont
  {Abe}} \emph {et~al.} (\bibinfo {collaboration} {Hyper-Kamiokande Working
  Group}),\ }\href@noop {} {\  (\bibinfo {year} {2014})},\ \Eprint
  {http://arxiv.org/abs/1412.4673} {arXiv:1412.4673 [physics.ins-det]}
  \BibitemShut {NoStop}%
\bibitem [{\citenamefont {Abe}\ \emph {et~al.}(2016)\citenamefont {Abe} \emph
  {et~al.}}]{T2HKK}%
  \BibitemOpen
  \bibfield  {author} {\bibinfo {author} {\bibfnamefont {K.}~\bibnamefont
  {Abe}} \emph {et~al.} (\bibinfo {collaboration} {Hyper-Kamiokande proto-}),\
  }\href@noop {} {\  (\bibinfo {year} {2016})},\ \Eprint
  {http://arxiv.org/abs/1611.06118} {arXiv:1611.06118 [hep-ex]} \BibitemShut
  {NoStop}%
\bibitem [{\citenamefont {Acciarri}\ \emph {et~al.}(2015)\citenamefont
  {Acciarri} \emph {et~al.}}]{Acciarri:2015uup}%
  \BibitemOpen
  \bibfield  {author} {\bibinfo {author} {\bibfnamefont {R.}~\bibnamefont
  {Acciarri}} \emph {et~al.} (\bibinfo {collaboration} {DUNE}),\ }\href@noop {}
  {\  (\bibinfo {year} {2015})},\ \Eprint {http://arxiv.org/abs/1512.06148}
  {arXiv:1512.06148 [physics.ins-det]} \BibitemShut {NoStop}%
\bibitem [{\citenamefont {Fukasawa}\ \emph
  {et~al.}(2016{\natexlab{a}})\citenamefont {Fukasawa}, \citenamefont {Ghosh},\
  and\ \citenamefont {Yasuda}}]{Fukasawa:2016yue}%
  \BibitemOpen
  \bibfield  {author} {\bibinfo {author} {\bibfnamefont {S.}~\bibnamefont
  {Fukasawa}}, \bibinfo {author} {\bibfnamefont {M.}~\bibnamefont {Ghosh}}, \
  and\ \bibinfo {author} {\bibfnamefont {O.}~\bibnamefont {Yasuda}},\
  }\href@noop {} {\  (\bibinfo {year} {2016}{\natexlab{a}})},\ \Eprint
  {http://arxiv.org/abs/1607.03758} {arXiv:1607.03758 [hep-ph]} \BibitemShut
  {NoStop}%
\bibitem [{\citenamefont {Ballett}\ \emph {et~al.}(2016)\citenamefont
  {Ballett}, \citenamefont {King}, \citenamefont {Pascoli}, \citenamefont
  {Prouse},\ and\ \citenamefont {Wang}}]{Ballett:2016daj}%
  \BibitemOpen
  \bibfield  {author} {\bibinfo {author} {\bibfnamefont {P.}~\bibnamefont
  {Ballett}}, \bibinfo {author} {\bibfnamefont {S.~F.}\ \bibnamefont {King}},
  \bibinfo {author} {\bibfnamefont {S.}~\bibnamefont {Pascoli}}, \bibinfo
  {author} {\bibfnamefont {N.~W.}\ \bibnamefont {Prouse}}, \ and\ \bibinfo
  {author} {\bibfnamefont {T.}~\bibnamefont {Wang}},\ }\href@noop {} {\
  (\bibinfo {year} {2016})},\ \Eprint {http://arxiv.org/abs/1612.07275}
  {arXiv:1612.07275 [hep-ph]} \BibitemShut {NoStop}%
\bibitem [{\citenamefont {Wolfenstein}(1978)}]{Wolfenstein:1977ue}%
  \BibitemOpen
  \bibfield  {author} {\bibinfo {author} {\bibfnamefont {L.}~\bibnamefont
  {Wolfenstein}},\ }\href {\doibase 10.1103/PhysRevD.17.2369} {\bibfield
  {journal} {\bibinfo  {journal} {Phys. Rev.}\ }\textbf {\bibinfo {volume}
  {D17}},\ \bibinfo {pages} {2369} (\bibinfo {year} {1978})}\BibitemShut
  {NoStop}%
\bibitem [{\citenamefont {Guzzo}\ \emph {et~al.}(1991)\citenamefont {Guzzo},
  \citenamefont {Masiero},\ and\ \citenamefont {Petcov}}]{Guzzo:1991hi}%
  \BibitemOpen
  \bibfield  {author} {\bibinfo {author} {\bibfnamefont {M.~M.}\ \bibnamefont
  {Guzzo}}, \bibinfo {author} {\bibfnamefont {A.}~\bibnamefont {Masiero}}, \
  and\ \bibinfo {author} {\bibfnamefont {S.~T.}\ \bibnamefont {Petcov}},\
  }\href {\doibase 10.1016/0370-2693(91)90984-X} {\bibfield  {journal}
  {\bibinfo  {journal} {Phys. Lett.}\ }\textbf {\bibinfo {volume} {B260}},\
  \bibinfo {pages} {154} (\bibinfo {year} {1991})}\BibitemShut {NoStop}%
\bibitem [{\citenamefont {Roulet}(1991)}]{Roulet:1991sm}%
  \BibitemOpen
  \bibfield  {author} {\bibinfo {author} {\bibfnamefont {E.}~\bibnamefont
  {Roulet}},\ }\href {\doibase 10.1103/PhysRevD.44.R935} {\bibfield  {journal}
  {\bibinfo  {journal} {Phys. Rev.}\ }\textbf {\bibinfo {volume} {D44}},\
  \bibinfo {pages} {R935} (\bibinfo {year} {1991})}\BibitemShut {NoStop}%
\bibitem [{\citenamefont {Ohlsson}(2013)}]{Ohlsson:2012kf}%
  \BibitemOpen
  \bibfield  {author} {\bibinfo {author} {\bibfnamefont {T.}~\bibnamefont
  {Ohlsson}},\ }\href {\doibase 10.1088/0034-4885/76/4/044201} {\bibfield
  {journal} {\bibinfo  {journal} {Rept. Prog. Phys.}\ }\textbf {\bibinfo
  {volume} {76}},\ \bibinfo {pages} {044201} (\bibinfo {year} {2013})},\
  \Eprint {http://arxiv.org/abs/1209.2710} {arXiv:1209.2710 [hep-ph]}
  \BibitemShut {NoStop}%
\bibitem [{\citenamefont {Miranda}\ and\ \citenamefont
  {Nunokawa}(2015)}]{Miranda:2015dra}%
  \BibitemOpen
  \bibfield  {author} {\bibinfo {author} {\bibfnamefont {O.~G.}\ \bibnamefont
  {Miranda}}\ and\ \bibinfo {author} {\bibfnamefont {H.}~\bibnamefont
  {Nunokawa}},\ }\href {\doibase 10.1088/1367-2630/17/9/095002} {\bibfield
  {journal} {\bibinfo  {journal} {New J. Phys.}\ }\textbf {\bibinfo {volume}
  {17}},\ \bibinfo {pages} {095002} (\bibinfo {year} {2015})},\ \Eprint
  {http://arxiv.org/abs/1505.06254} {arXiv:1505.06254 [hep-ph]} \BibitemShut
  {NoStop}%
\bibitem [{\citenamefont {Gonzalez-Garcia}\ and\ \citenamefont
  {Maltoni}(2013)}]{Gonzalez-Garcia:2013usa}%
  \BibitemOpen
  \bibfield  {author} {\bibinfo {author} {\bibfnamefont {M.~C.}\ \bibnamefont
  {Gonzalez-Garcia}}\ and\ \bibinfo {author} {\bibfnamefont {M.}~\bibnamefont
  {Maltoni}},\ }\href {\doibase 10.1007/JHEP09(2013)152} {\bibfield  {journal}
  {\bibinfo  {journal} {JHEP}\ }\textbf {\bibinfo {volume} {09}},\ \bibinfo
  {pages} {152} (\bibinfo {year} {2013})},\ \Eprint
  {http://arxiv.org/abs/1307.3092} {arXiv:1307.3092 [hep-ph]} \BibitemShut
  {NoStop}%
\bibitem [{\citenamefont {Friedland}\ and\ \citenamefont
  {Shoemaker}(2012)}]{Friedland:2012tq}%
  \BibitemOpen
  \bibfield  {author} {\bibinfo {author} {\bibfnamefont {A.}~\bibnamefont
  {Friedland}}\ and\ \bibinfo {author} {\bibfnamefont {I.~M.}\ \bibnamefont
  {Shoemaker}},\ }\href@noop {} {\  (\bibinfo {year} {2012})},\ \Eprint
  {http://arxiv.org/abs/1207.6642} {arXiv:1207.6642 [hep-ph]} \BibitemShut
  {NoStop}%
\bibitem [{\citenamefont {Adhikari}\ \emph {et~al.}(2012)\citenamefont
  {Adhikari}, \citenamefont {Chakraborty}, \citenamefont {Dasgupta},\ and\
  \citenamefont {Roy}}]{Adhikari:2012vc}%
  \BibitemOpen
  \bibfield  {author} {\bibinfo {author} {\bibfnamefont {R.}~\bibnamefont
  {Adhikari}}, \bibinfo {author} {\bibfnamefont {S.}~\bibnamefont
  {Chakraborty}}, \bibinfo {author} {\bibfnamefont {A.}~\bibnamefont
  {Dasgupta}}, \ and\ \bibinfo {author} {\bibfnamefont {S.}~\bibnamefont
  {Roy}},\ }\href {\doibase 10.1103/PhysRevD.86.073010} {\bibfield  {journal}
  {\bibinfo  {journal} {Phys. Rev.}\ }\textbf {\bibinfo {volume} {D86}},\
  \bibinfo {pages} {073010} (\bibinfo {year} {2012})},\ \Eprint
  {http://arxiv.org/abs/1201.3047} {arXiv:1201.3047 [hep-ph]} \BibitemShut
  {NoStop}%
\bibitem [{\citenamefont {Masud}\ \emph {et~al.}(2016)\citenamefont {Masud},
  \citenamefont {Chatterjee},\ and\ \citenamefont {Mehta}}]{Masud:2015xva}%
  \BibitemOpen
  \bibfield  {author} {\bibinfo {author} {\bibfnamefont {M.}~\bibnamefont
  {Masud}}, \bibinfo {author} {\bibfnamefont {A.}~\bibnamefont {Chatterjee}}, \
  and\ \bibinfo {author} {\bibfnamefont {P.}~\bibnamefont {Mehta}},\ }\href
  {\doibase 10.1088/0954-3899/43/9/095005/meta, 10.1088/0954-3899/43/9/095005}
  {\bibfield  {journal} {\bibinfo  {journal} {J. Phys.}\ }\textbf {\bibinfo
  {volume} {G43}},\ \bibinfo {pages} {095005} (\bibinfo {year} {2016})},\
  \Eprint {http://arxiv.org/abs/1510.08261} {arXiv:1510.08261 [hep-ph]}
  \BibitemShut {NoStop}%
\bibitem [{\citenamefont {de~Gouvêa}\ and\ \citenamefont
  {Kelly}(2016)}]{deGouvea:2015ndi}%
  \BibitemOpen
  \bibfield  {author} {\bibinfo {author} {\bibfnamefont {A.}~\bibnamefont
  {de~Gouvêa}}\ and\ \bibinfo {author} {\bibfnamefont {K.~J.}\ \bibnamefont
  {Kelly}},\ }\href {\doibase 10.1016/j.nuclphysb.2016.03.013} {\bibfield
  {journal} {\bibinfo  {journal} {Nucl. Phys.}\ }\textbf {\bibinfo {volume}
  {B908}},\ \bibinfo {pages} {318} (\bibinfo {year} {2016})},\ \Eprint
  {http://arxiv.org/abs/1511.05562} {arXiv:1511.05562 [hep-ph]} \BibitemShut
  {NoStop}%
\bibitem [{\citenamefont {Rahman}\ \emph {et~al.}(2015)\citenamefont {Rahman},
  \citenamefont {Dasgupta},\ and\ \citenamefont {Adhikari}}]{Rahman:2015vqa}%
  \BibitemOpen
  \bibfield  {author} {\bibinfo {author} {\bibfnamefont {Z.}~\bibnamefont
  {Rahman}}, \bibinfo {author} {\bibfnamefont {A.}~\bibnamefont {Dasgupta}}, \
  and\ \bibinfo {author} {\bibfnamefont {R.}~\bibnamefont {Adhikari}},\ }\href
  {\doibase 10.1088/0954-3899/42/6/065001} {\bibfield  {journal} {\bibinfo
  {journal} {J. Phys.}\ }\textbf {\bibinfo {volume} {G42}},\ \bibinfo {pages}
  {065001} (\bibinfo {year} {2015})},\ \Eprint
  {http://arxiv.org/abs/1503.03248} {arXiv:1503.03248 [hep-ph]} \BibitemShut
  {NoStop}%
\bibitem [{\citenamefont {Coloma}(2016)}]{Coloma:2015kiu}%
  \BibitemOpen
  \bibfield  {author} {\bibinfo {author} {\bibfnamefont {P.}~\bibnamefont
  {Coloma}},\ }\href {\doibase 10.1007/JHEP03(2016)016} {\bibfield  {journal}
  {\bibinfo  {journal} {JHEP}\ }\textbf {\bibinfo {volume} {03}},\ \bibinfo
  {pages} {016} (\bibinfo {year} {2016})},\ \Eprint
  {http://arxiv.org/abs/1511.06357} {arXiv:1511.06357 [hep-ph]} \BibitemShut
  {NoStop}%
\bibitem [{\citenamefont {Liao}\ \emph
  {et~al.}(2016{\natexlab{a}})\citenamefont {Liao}, \citenamefont {Marfatia},\
  and\ \citenamefont {Whisnant}}]{Liao:2016hsa}%
  \BibitemOpen
  \bibfield  {author} {\bibinfo {author} {\bibfnamefont {J.}~\bibnamefont
  {Liao}}, \bibinfo {author} {\bibfnamefont {D.}~\bibnamefont {Marfatia}}, \
  and\ \bibinfo {author} {\bibfnamefont {K.}~\bibnamefont {Whisnant}},\ }\href
  {\doibase 10.1103/PhysRevD.93.093016} {\bibfield  {journal} {\bibinfo
  {journal} {Phys. Rev.}\ }\textbf {\bibinfo {volume} {D93}},\ \bibinfo {pages}
  {093016} (\bibinfo {year} {2016}{\natexlab{a}})},\ \Eprint
  {http://arxiv.org/abs/1601.00927} {arXiv:1601.00927 [hep-ph]} \BibitemShut
  {NoStop}%
\bibitem [{\citenamefont {Soumya}\ and\ \citenamefont
  {Mohanta}(2016{\natexlab{b}})}]{Soumya:2016enw}%
  \BibitemOpen
  \bibfield  {author} {\bibinfo {author} {\bibfnamefont {C.}~\bibnamefont
  {Soumya}}\ and\ \bibinfo {author} {\bibfnamefont {R.}~\bibnamefont
  {Mohanta}},\ }\href {\doibase 10.1103/PhysRevD.94.053008} {\bibfield
  {journal} {\bibinfo  {journal} {Phys. Rev.}\ }\textbf {\bibinfo {volume}
  {D94}},\ \bibinfo {pages} {053008} (\bibinfo {year} {2016}{\natexlab{b}})},\
  \Eprint {http://arxiv.org/abs/1603.02184} {arXiv:1603.02184 [hep-ph]}
  \BibitemShut {NoStop}%
\bibitem [{\citenamefont {Blennow}\ \emph
  {et~al.}(2016{\natexlab{a}})\citenamefont {Blennow}, \citenamefont {Choubey},
  \citenamefont {Ohlsson}, \citenamefont {Pramanik},\ and\ \citenamefont
  {Raut}}]{Blennow:2016etl}%
  \BibitemOpen
  \bibfield  {author} {\bibinfo {author} {\bibfnamefont {M.}~\bibnamefont
  {Blennow}}, \bibinfo {author} {\bibfnamefont {S.}~\bibnamefont {Choubey}},
  \bibinfo {author} {\bibfnamefont {T.}~\bibnamefont {Ohlsson}}, \bibinfo
  {author} {\bibfnamefont {D.}~\bibnamefont {Pramanik}}, \ and\ \bibinfo
  {author} {\bibfnamefont {S.~K.}\ \bibnamefont {Raut}},\ }\href {\doibase
  10.1007/JHEP08(2016)090} {\bibfield  {journal} {\bibinfo  {journal} {JHEP}\
  }\textbf {\bibinfo {volume} {08}},\ \bibinfo {pages} {090} (\bibinfo {year}
  {2016}{\natexlab{a}})},\ \Eprint {http://arxiv.org/abs/1606.08851}
  {arXiv:1606.08851 [hep-ph]} \BibitemShut {NoStop}%
\bibitem [{\citenamefont {Forero}\ and\ \citenamefont
  {Huber}(2016)}]{Forero:2016cmb}%
  \BibitemOpen
  \bibfield  {author} {\bibinfo {author} {\bibfnamefont {D.~V.}\ \bibnamefont
  {Forero}}\ and\ \bibinfo {author} {\bibfnamefont {P.}~\bibnamefont {Huber}},\
  }\href {\doibase 10.1103/PhysRevLett.117.031801} {\bibfield  {journal}
  {\bibinfo  {journal} {Phys. Rev. Lett.}\ }\textbf {\bibinfo {volume} {117}},\
  \bibinfo {pages} {031801} (\bibinfo {year} {2016})},\ \Eprint
  {http://arxiv.org/abs/1601.03736} {arXiv:1601.03736 [hep-ph]} \BibitemShut
  {NoStop}%
\bibitem [{\citenamefont {Huitu}\ \emph {et~al.}(2016)\citenamefont {Huitu},
  \citenamefont {Kärkkäinen}, \citenamefont {Maalampi},\ and\ \citenamefont
  {Vihonen}}]{Huitu:2016bmb}%
  \BibitemOpen
  \bibfield  {author} {\bibinfo {author} {\bibfnamefont {K.}~\bibnamefont
  {Huitu}}, \bibinfo {author} {\bibfnamefont {T.~J.}\ \bibnamefont
  {Kärkkäinen}}, \bibinfo {author} {\bibfnamefont {J.}~\bibnamefont
  {Maalampi}}, \ and\ \bibinfo {author} {\bibfnamefont {S.}~\bibnamefont
  {Vihonen}},\ }\href {\doibase 10.1103/PhysRevD.93.053016} {\bibfield
  {journal} {\bibinfo  {journal} {Phys. Rev.}\ }\textbf {\bibinfo {volume}
  {D93}},\ \bibinfo {pages} {053016} (\bibinfo {year} {2016})},\ \Eprint
  {http://arxiv.org/abs/1601.07730} {arXiv:1601.07730 [hep-ph]} \BibitemShut
  {NoStop}%
\bibitem [{\citenamefont {Bakhti}\ and\ \citenamefont
  {Farzan}(2016)}]{Bakhti:2016prn}%
  \BibitemOpen
  \bibfield  {author} {\bibinfo {author} {\bibfnamefont {P.}~\bibnamefont
  {Bakhti}}\ and\ \bibinfo {author} {\bibfnamefont {Y.}~\bibnamefont
  {Farzan}},\ }\href {\doibase 10.1007/JHEP07(2016)109} {\bibfield  {journal}
  {\bibinfo  {journal} {JHEP}\ }\textbf {\bibinfo {volume} {07}},\ \bibinfo
  {pages} {109} (\bibinfo {year} {2016})},\ \Eprint
  {http://arxiv.org/abs/1602.07099} {arXiv:1602.07099 [hep-ph]} \BibitemShut
  {NoStop}%
\bibitem [{\citenamefont {Masud}\ and\ \citenamefont
  {Mehta}(2016{\natexlab{a}})}]{Masud:2016bvp}%
  \BibitemOpen
  \bibfield  {author} {\bibinfo {author} {\bibfnamefont {M.}~\bibnamefont
  {Masud}}\ and\ \bibinfo {author} {\bibfnamefont {P.}~\bibnamefont {Mehta}},\
  }\href {\doibase 10.1103/PhysRevD.94.013014} {\bibfield  {journal} {\bibinfo
  {journal} {Phys. Rev.}\ }\textbf {\bibinfo {volume} {D94}},\ \bibinfo {pages}
  {013014} (\bibinfo {year} {2016}{\natexlab{a}})},\ \Eprint
  {http://arxiv.org/abs/1603.01380} {arXiv:1603.01380 [hep-ph]} \BibitemShut
  {NoStop}%
\bibitem [{\citenamefont {Coloma}\ and\ \citenamefont
  {Schwetz}(2016)}]{Coloma:2016gei}%
  \BibitemOpen
  \bibfield  {author} {\bibinfo {author} {\bibfnamefont {P.}~\bibnamefont
  {Coloma}}\ and\ \bibinfo {author} {\bibfnamefont {T.}~\bibnamefont
  {Schwetz}},\ }\href {\doibase 10.1103/PhysRevD.94.055005} {\bibfield
  {journal} {\bibinfo  {journal} {Phys. Rev.}\ }\textbf {\bibinfo {volume}
  {D94}},\ \bibinfo {pages} {055005} (\bibinfo {year} {2016})},\ \Eprint
  {http://arxiv.org/abs/1604.05772} {arXiv:1604.05772 [hep-ph]} \BibitemShut
  {NoStop}%
\bibitem [{\citenamefont {Masud}\ and\ \citenamefont
  {Mehta}(2016{\natexlab{b}})}]{Masud:2016gcl}%
  \BibitemOpen
  \bibfield  {author} {\bibinfo {author} {\bibfnamefont {M.}~\bibnamefont
  {Masud}}\ and\ \bibinfo {author} {\bibfnamefont {P.}~\bibnamefont {Mehta}},\
  }\href {\doibase 10.1103/PhysRevD.94.053007} {\bibfield  {journal} {\bibinfo
  {journal} {Phys. Rev.}\ }\textbf {\bibinfo {volume} {D94}},\ \bibinfo {pages}
  {053007} (\bibinfo {year} {2016}{\natexlab{b}})},\ \Eprint
  {http://arxiv.org/abs/1606.05662} {arXiv:1606.05662 [hep-ph]} \BibitemShut
  {NoStop}%
\bibitem [{\citenamefont {Agarwalla}\ \emph {et~al.}(2016)\citenamefont
  {Agarwalla}, \citenamefont {Chatterjee},\ and\ \citenamefont
  {Palazzo}}]{Agarwalla:2016fkh}%
  \BibitemOpen
  \bibfield  {author} {\bibinfo {author} {\bibfnamefont {S.~K.}\ \bibnamefont
  {Agarwalla}}, \bibinfo {author} {\bibfnamefont {S.~S.}\ \bibnamefont
  {Chatterjee}}, \ and\ \bibinfo {author} {\bibfnamefont {A.}~\bibnamefont
  {Palazzo}},\ }\href {\doibase 10.1016/j.physletb.2016.09.020} {\bibfield
  {journal} {\bibinfo  {journal} {Phys. Lett.}\ }\textbf {\bibinfo {volume}
  {B762}},\ \bibinfo {pages} {64} (\bibinfo {year} {2016})},\ \Eprint
  {http://arxiv.org/abs/1607.01745} {arXiv:1607.01745 [hep-ph]} \BibitemShut
  {NoStop}%
\bibitem [{\citenamefont {Ge}\ and\ \citenamefont
  {Smirnov}(2016)}]{Ge:2016dlx}%
  \BibitemOpen
  \bibfield  {author} {\bibinfo {author} {\bibfnamefont {S.-F.}\ \bibnamefont
  {Ge}}\ and\ \bibinfo {author} {\bibfnamefont {A.~{\relax Yu}.}\ \bibnamefont
  {Smirnov}},\ }\href {\doibase 10.1007/JHEP10(2016)138} {\bibfield  {journal}
  {\bibinfo  {journal} {JHEP}\ }\textbf {\bibinfo {volume} {10}},\ \bibinfo
  {pages} {138} (\bibinfo {year} {2016})},\ \Eprint
  {http://arxiv.org/abs/1607.08513} {arXiv:1607.08513 [hep-ph]} \BibitemShut
  {NoStop}%
\bibitem [{\citenamefont {Liao}\ \emph
  {et~al.}(2016{\natexlab{b}})\citenamefont {Liao}, \citenamefont {Marfatia},\
  and\ \citenamefont {Whisnant}}]{Liao:2016bgf}%
  \BibitemOpen
  \bibfield  {author} {\bibinfo {author} {\bibfnamefont {J.}~\bibnamefont
  {Liao}}, \bibinfo {author} {\bibfnamefont {D.}~\bibnamefont {Marfatia}}, \
  and\ \bibinfo {author} {\bibfnamefont {K.}~\bibnamefont {Whisnant}},\
  }\href@noop {} {\  (\bibinfo {year} {2016}{\natexlab{b}})},\ \Eprint
  {http://arxiv.org/abs/1609.01786} {arXiv:1609.01786 [hep-ph]} \BibitemShut
  {NoStop}%
\bibitem [{\citenamefont {Fukasawa}\ \emph
  {et~al.}(2016{\natexlab{b}})\citenamefont {Fukasawa}, \citenamefont {Ghosh},\
  and\ \citenamefont {Yasuda}}]{Fukasawa:2016gvm}%
  \BibitemOpen
  \bibfield  {author} {\bibinfo {author} {\bibfnamefont {S.}~\bibnamefont
  {Fukasawa}}, \bibinfo {author} {\bibfnamefont {M.}~\bibnamefont {Ghosh}}, \
  and\ \bibinfo {author} {\bibfnamefont {O.}~\bibnamefont {Yasuda}},\
  }\href@noop {} {\  (\bibinfo {year} {2016}{\natexlab{b}})},\ \Eprint
  {http://arxiv.org/abs/1609.04204} {arXiv:1609.04204 [hep-ph]} \BibitemShut
  {NoStop}%
\bibitem [{\citenamefont {Blennow}\ \emph
  {et~al.}(2016{\natexlab{b}})\citenamefont {Blennow}, \citenamefont {Coloma},
  \citenamefont {Fernandez-Martinez}, \citenamefont {Hernandez-Garcia},\ and\
  \citenamefont {Lopez-Pavon}}]{Blennow:2016jkn}%
  \BibitemOpen
  \bibfield  {author} {\bibinfo {author} {\bibfnamefont {M.}~\bibnamefont
  {Blennow}}, \bibinfo {author} {\bibfnamefont {P.}~\bibnamefont {Coloma}},
  \bibinfo {author} {\bibfnamefont {E.}~\bibnamefont {Fernandez-Martinez}},
  \bibinfo {author} {\bibfnamefont {J.}~\bibnamefont {Hernandez-Garcia}}, \
  and\ \bibinfo {author} {\bibfnamefont {J.}~\bibnamefont {Lopez-Pavon}},\
  }\href@noop {} {\  (\bibinfo {year} {2016}{\natexlab{b}})},\ \Eprint
  {http://arxiv.org/abs/1609.08637} {arXiv:1609.08637 [hep-ph]} \BibitemShut
  {NoStop}%
\bibitem [{\citenamefont {Liao}\ \emph {et~al.}(2017)\citenamefont {Liao},
  \citenamefont {Marfatia},\ and\ \citenamefont {Whisnant}}]{Liao:2016orc}%
  \BibitemOpen
  \bibfield  {author} {\bibinfo {author} {\bibfnamefont {J.}~\bibnamefont
  {Liao}}, \bibinfo {author} {\bibfnamefont {D.}~\bibnamefont {Marfatia}}, \
  and\ \bibinfo {author} {\bibfnamefont {K.}~\bibnamefont {Whisnant}},\ }\href
  {\doibase 10.1007/JHEP01(2017)071} {\bibfield  {journal} {\bibinfo  {journal}
  {JHEP}\ }\textbf {\bibinfo {volume} {01}},\ \bibinfo {pages} {071} (\bibinfo
  {year} {2017})},\ \Eprint {http://arxiv.org/abs/1612.01443} {arXiv:1612.01443
  [hep-ph]} \BibitemShut {NoStop}%
\bibitem [{\citenamefont {Deepthi}\ \emph {et~al.}(2016)\citenamefont
  {Deepthi}, \citenamefont {Goswami},\ and\ \citenamefont
  {Nath}}]{Deepthi:2016erc}%
  \BibitemOpen
  \bibfield  {author} {\bibinfo {author} {\bibfnamefont {K.~N.}\ \bibnamefont
  {Deepthi}}, \bibinfo {author} {\bibfnamefont {S.}~\bibnamefont {Goswami}}, \
  and\ \bibinfo {author} {\bibfnamefont {N.}~\bibnamefont {Nath}},\ }\href@noop
  {} {\  (\bibinfo {year} {2016})},\ \Eprint {http://arxiv.org/abs/1612.00784}
  {arXiv:1612.00784 [hep-ph]} \BibitemShut {NoStop}%
\bibitem [{\citenamefont {Fukasawa}\ \emph {et~al.}(2017)\citenamefont
  {Fukasawa}, \citenamefont {Ghosh},\ and\ \citenamefont
  {Yasuda}}]{Fukasawa:2016lew}%
  \BibitemOpen
  \bibfield  {author} {\bibinfo {author} {\bibfnamefont {S.}~\bibnamefont
  {Fukasawa}}, \bibinfo {author} {\bibfnamefont {M.}~\bibnamefont {Ghosh}}, \
  and\ \bibinfo {author} {\bibfnamefont {O.}~\bibnamefont {Yasuda}},\ }\href
  {\doibase 10.1103/PhysRevD.95.055005} {\bibfield  {journal} {\bibinfo
  {journal} {Phys. Rev.}\ }\textbf {\bibinfo {volume} {D95}},\ \bibinfo {pages}
  {055005} (\bibinfo {year} {2017})},\ \Eprint
  {http://arxiv.org/abs/1611.06141} {arXiv:1611.06141 [hep-ph]} \BibitemShut
  {NoStop}%
\bibitem [{\citenamefont {Huber}\ \emph {et~al.}(2002)\citenamefont {Huber},
  \citenamefont {Lindner},\ and\ \citenamefont {Winter}}]{Huber:2002mx}%
  \BibitemOpen
  \bibfield  {author} {\bibinfo {author} {\bibfnamefont {P.}~\bibnamefont
  {Huber}}, \bibinfo {author} {\bibfnamefont {M.}~\bibnamefont {Lindner}}, \
  and\ \bibinfo {author} {\bibfnamefont {W.}~\bibnamefont {Winter}},\ }\href
  {\doibase 10.1016/S0550-3213(02)00825-8} {\bibfield  {journal} {\bibinfo
  {journal} {Nucl. Phys.}\ }\textbf {\bibinfo {volume} {B645}},\ \bibinfo
  {pages} {3} (\bibinfo {year} {2002})},\ \Eprint
  {http://arxiv.org/abs/hep-ph/0204352} {arXiv:hep-ph/0204352 [hep-ph]}
  \BibitemShut {NoStop}%
\bibitem [{\citenamefont {Ohlsson}\ and\ \citenamefont
  {Winter}(2003)}]{Ohlsson:2003ip}%
  \BibitemOpen
  \bibfield  {author} {\bibinfo {author} {\bibfnamefont {T.}~\bibnamefont
  {Ohlsson}}\ and\ \bibinfo {author} {\bibfnamefont {W.}~\bibnamefont
  {Winter}},\ }\href {\doibase 10.1103/PhysRevD.68.073007} {\bibfield
  {journal} {\bibinfo  {journal} {Phys. Rev.}\ }\textbf {\bibinfo {volume}
  {D68}},\ \bibinfo {pages} {073007} (\bibinfo {year} {2003})},\ \Eprint
  {http://arxiv.org/abs/hep-ph/0307178} {arXiv:hep-ph/0307178 [hep-ph]}
  \BibitemShut {NoStop}%
\bibitem [{\citenamefont {Barger}\ \emph {et~al.}(2007)\citenamefont {Barger},
  \citenamefont {Huber}, \citenamefont {Marfatia},\ and\ \citenamefont
  {Winter}}]{Barger:2007jq}%
  \BibitemOpen
  \bibfield  {author} {\bibinfo {author} {\bibfnamefont {V.}~\bibnamefont
  {Barger}}, \bibinfo {author} {\bibfnamefont {P.}~\bibnamefont {Huber}},
  \bibinfo {author} {\bibfnamefont {D.}~\bibnamefont {Marfatia}}, \ and\
  \bibinfo {author} {\bibfnamefont {W.}~\bibnamefont {Winter}},\ }\href
  {\doibase 10.1103/PhysRevD.76.053005} {\bibfield  {journal} {\bibinfo
  {journal} {Phys. Rev.}\ }\textbf {\bibinfo {volume} {D76}},\ \bibinfo {pages}
  {053005} (\bibinfo {year} {2007})},\ \Eprint
  {http://arxiv.org/abs/hep-ph/0703029} {arXiv:hep-ph/0703029 [hep-ph]}
  \BibitemShut {NoStop}%
\bibitem [{\citenamefont {Huber}\ \emph {et~al.}(2008)\citenamefont {Huber},
  \citenamefont {Mezzetto},\ and\ \citenamefont {Schwetz}}]{Huber:2007em}%
  \BibitemOpen
  \bibfield  {author} {\bibinfo {author} {\bibfnamefont {P.}~\bibnamefont
  {Huber}}, \bibinfo {author} {\bibfnamefont {M.}~\bibnamefont {Mezzetto}}, \
  and\ \bibinfo {author} {\bibfnamefont {T.}~\bibnamefont {Schwetz}},\ }\href
  {\doibase 10.1088/1126-6708/2008/03/021} {\bibfield  {journal} {\bibinfo
  {journal} {JHEP}\ }\textbf {\bibinfo {volume} {03}},\ \bibinfo {pages} {021}
  (\bibinfo {year} {2008})},\ \Eprint {http://arxiv.org/abs/0711.2950}
  {arXiv:0711.2950 [hep-ph]} \BibitemShut {NoStop}%
\bibitem [{\citenamefont {Coloma}\ and\ \citenamefont
  {Fernandez-Martinez}(2012)}]{Coloma:2011pg}%
  \BibitemOpen
  \bibfield  {author} {\bibinfo {author} {\bibfnamefont {P.}~\bibnamefont
  {Coloma}}\ and\ \bibinfo {author} {\bibfnamefont {E.}~\bibnamefont
  {Fernandez-Martinez}},\ }\href {\doibase 10.1007/JHEP04(2012)089} {\bibfield
  {journal} {\bibinfo  {journal} {JHEP}\ }\textbf {\bibinfo {volume} {04}},\
  \bibinfo {pages} {089} (\bibinfo {year} {2012})},\ \Eprint
  {http://arxiv.org/abs/1110.4583} {arXiv:1110.4583 [hep-ph]} \BibitemShut
  {NoStop}%
\bibitem [{\citenamefont {Tang}\ and\ \citenamefont
  {Winter}(2009)}]{Tang:2009na}%
  \BibitemOpen
  \bibfield  {author} {\bibinfo {author} {\bibfnamefont {J.}~\bibnamefont
  {Tang}}\ and\ \bibinfo {author} {\bibfnamefont {W.}~\bibnamefont {Winter}},\
  }\href {\doibase 10.1103/PhysRevD.80.053001} {\bibfield  {journal} {\bibinfo
  {journal} {Phys. Rev.}\ }\textbf {\bibinfo {volume} {D80}},\ \bibinfo {pages}
  {053001} (\bibinfo {year} {2009})},\ \Eprint {http://arxiv.org/abs/0903.3039}
  {arXiv:0903.3039 [hep-ph]} \BibitemShut {NoStop}%
\bibitem [{\citenamefont {Ankowski}\ and\ \citenamefont
  {Mariani}(2016)}]{Ankowski:2016jdd}%
  \BibitemOpen
  \bibfield  {author} {\bibinfo {author} {\bibfnamefont {A.~M.}\ \bibnamefont
  {Ankowski}}\ and\ \bibinfo {author} {\bibfnamefont {C.}~\bibnamefont
  {Mariani}},\ }\href@noop {} {\  (\bibinfo {year} {2016})},\ \Eprint
  {http://arxiv.org/abs/1609.00258} {arXiv:1609.00258 [hep-ph]} \BibitemShut
  {NoStop}%
\bibitem [{\citenamefont {Coloma}\ \emph {et~al.}(2013)\citenamefont {Coloma},
  \citenamefont {Huber}, \citenamefont {Kopp},\ and\ \citenamefont
  {Winter}}]{Coloma:2012ji}%
  \BibitemOpen
  \bibfield  {author} {\bibinfo {author} {\bibfnamefont {P.}~\bibnamefont
  {Coloma}}, \bibinfo {author} {\bibfnamefont {P.}~\bibnamefont {Huber}},
  \bibinfo {author} {\bibfnamefont {J.}~\bibnamefont {Kopp}}, \ and\ \bibinfo
  {author} {\bibfnamefont {W.}~\bibnamefont {Winter}},\ }\href {\doibase
  10.1103/PhysRevD.87.033004} {\bibfield  {journal} {\bibinfo  {journal} {Phys.
  Rev.}\ }\textbf {\bibinfo {volume} {D87}},\ \bibinfo {pages} {033004}
  (\bibinfo {year} {2013})},\ \Eprint {http://arxiv.org/abs/1209.5973}
  {arXiv:1209.5973 [hep-ph]} \BibitemShut {NoStop}%
\bibitem [{\citenamefont {Huber}\ \emph {et~al.}(2005)\citenamefont {Huber},
  \citenamefont {Lindner},\ and\ \citenamefont {Winter}}]{Huber:2004ka}%
  \BibitemOpen
  \bibfield  {author} {\bibinfo {author} {\bibfnamefont {P.}~\bibnamefont
  {Huber}}, \bibinfo {author} {\bibfnamefont {M.}~\bibnamefont {Lindner}}, \
  and\ \bibinfo {author} {\bibfnamefont {W.}~\bibnamefont {Winter}},\ }\href
  {\doibase 10.1016/j.cpc.2005.01.003} {\bibfield  {journal} {\bibinfo
  {journal} {Comput. Phys. Commun.}\ }\textbf {\bibinfo {volume} {167}},\
  \bibinfo {pages} {195} (\bibinfo {year} {2005})},\ \Eprint
  {http://arxiv.org/abs/hep-ph/0407333} {arXiv:hep-ph/0407333} \BibitemShut
  {NoStop}%
\bibitem [{\citenamefont {Huber}\ \emph {et~al.}(2007)\citenamefont {Huber},
  \citenamefont {Kopp}, \citenamefont {Lindner}, \citenamefont {Rolinec},\ and\
  \citenamefont {Winter}}]{Huber:2007ji}%
  \BibitemOpen
  \bibfield  {author} {\bibinfo {author} {\bibfnamefont {P.}~\bibnamefont
  {Huber}}, \bibinfo {author} {\bibfnamefont {J.}~\bibnamefont {Kopp}},
  \bibinfo {author} {\bibfnamefont {M.}~\bibnamefont {Lindner}}, \bibinfo
  {author} {\bibfnamefont {M.}~\bibnamefont {Rolinec}}, \ and\ \bibinfo
  {author} {\bibfnamefont {W.}~\bibnamefont {Winter}},\ }\href {\doibase
  10.1016/j.cpc.2007.05.004} {\bibfield  {journal} {\bibinfo  {journal}
  {Comput. Phys. Commun.}\ }\textbf {\bibinfo {volume} {177}},\ \bibinfo
  {pages} {432} (\bibinfo {year} {2007})},\ \Eprint
  {http://arxiv.org/abs/hep-ph/0701187} {arXiv:hep-ph/0701187} \BibitemShut
  {NoStop}%
\bibitem [{\citenamefont {Blennow}\ and\ \citenamefont
  {Fernandez-Martinez}(2010)}]{Blennow:2009pk}%
  \BibitemOpen
  \bibfield  {author} {\bibinfo {author} {\bibfnamefont {M.}~\bibnamefont
  {Blennow}}\ and\ \bibinfo {author} {\bibfnamefont {E.}~\bibnamefont
  {Fernandez-Martinez}},\ }\href {\doibase 10.1016/j.cpc.2009.09.014}
  {\bibfield  {journal} {\bibinfo  {journal} {Comput. Phys. Commun.}\ }\textbf
  {\bibinfo {volume} {181}},\ \bibinfo {pages} {227} (\bibinfo {year}
  {2010})},\ \Eprint {http://arxiv.org/abs/0903.3985} {arXiv:0903.3985
  [hep-ph]} \BibitemShut {NoStop}%
\bibitem [{\citenamefont {Forero}\ \emph {et~al.}(2014)\citenamefont {Forero},
  \citenamefont {Tortola},\ and\ \citenamefont {Valle}}]{Forero:2014bxa}%
  \BibitemOpen
  \bibfield  {author} {\bibinfo {author} {\bibfnamefont {D.~V.}\ \bibnamefont
  {Forero}}, \bibinfo {author} {\bibfnamefont {M.}~\bibnamefont {Tortola}}, \
  and\ \bibinfo {author} {\bibfnamefont {J.~W.~F.}\ \bibnamefont {Valle}},\
  }\href {\doibase 10.1103/PhysRevD.90.093006} {\bibfield  {journal} {\bibinfo
  {journal} {Phys. Rev.}\ }\textbf {\bibinfo {volume} {D90}},\ \bibinfo {pages}
  {093006} (\bibinfo {year} {2014})},\ \Eprint {http://arxiv.org/abs/1405.7540}
  {arXiv:1405.7540 [hep-ph]} \BibitemShut {NoStop}%
\bibitem [{\citenamefont {Esteban}\ \emph {et~al.}(2016)\citenamefont
  {Esteban}, \citenamefont {Gonzalez-Garcia}, \citenamefont {Maltoni},
  \citenamefont {Martinez-Soler},\ and\ \citenamefont
  {Schwetz}}]{Esteban:2016qun}%
  \BibitemOpen
  \bibfield  {author} {\bibinfo {author} {\bibfnamefont {I.}~\bibnamefont
  {Esteban}}, \bibinfo {author} {\bibfnamefont {M.~C.}\ \bibnamefont
  {Gonzalez-Garcia}}, \bibinfo {author} {\bibfnamefont {M.}~\bibnamefont
  {Maltoni}}, \bibinfo {author} {\bibfnamefont {I.}~\bibnamefont
  {Martinez-Soler}}, \ and\ \bibinfo {author} {\bibfnamefont {T.}~\bibnamefont
  {Schwetz}},\ }\href@noop {} {\  (\bibinfo {year} {2016})},\ \Eprint
  {http://arxiv.org/abs/1611.01514} {arXiv:1611.01514 [hep-ph]} \BibitemShut
  {NoStop}%
\bibitem [{\citenamefont {Capozzi}\ \emph {et~al.}(2014)\citenamefont
  {Capozzi}, \citenamefont {Fogli}, \citenamefont {Lisi}, \citenamefont
  {Marrone}, \citenamefont {Montanino},\ and\ \citenamefont
  {Palazzo}}]{Capozzi:2013csa}%
  \BibitemOpen
  \bibfield  {author} {\bibinfo {author} {\bibfnamefont {F.}~\bibnamefont
  {Capozzi}}, \bibinfo {author} {\bibfnamefont {G.~L.}\ \bibnamefont {Fogli}},
  \bibinfo {author} {\bibfnamefont {E.}~\bibnamefont {Lisi}}, \bibinfo {author}
  {\bibfnamefont {A.}~\bibnamefont {Marrone}}, \bibinfo {author} {\bibfnamefont
  {D.}~\bibnamefont {Montanino}}, \ and\ \bibinfo {author} {\bibfnamefont
  {A.}~\bibnamefont {Palazzo}},\ }\href {\doibase 10.1103/PhysRevD.89.093018}
  {\bibfield  {journal} {\bibinfo  {journal} {Phys. Rev.}\ }\textbf {\bibinfo
  {volume} {D89}},\ \bibinfo {pages} {093018} (\bibinfo {year} {2014})},\
  \Eprint {http://arxiv.org/abs/1312.2878} {arXiv:1312.2878 [hep-ph]}
  \BibitemShut {NoStop}%
\bibitem [{\citenamefont {Nunokawa}\ \emph {et~al.}(2005)\citenamefont
  {Nunokawa}, \citenamefont {Parke},\ and\ \citenamefont {{Zukanovich
  Funchal}}}]{Nunokawa:2005nx}%
  \BibitemOpen
  \bibfield  {author} {\bibinfo {author} {\bibfnamefont {H.}~\bibnamefont
  {Nunokawa}}, \bibinfo {author} {\bibfnamefont {S.~J.}\ \bibnamefont {Parke}},
  \ and\ \bibinfo {author} {\bibfnamefont {R.}~\bibnamefont {{Zukanovich
  Funchal}}},\ }\href {\doibase 10.1103/PhysRevD.72.013009} {\bibfield
  {journal} {\bibinfo  {journal} {Phys.Rev.}\ }\textbf {\bibinfo {volume}
  {D72}},\ \bibinfo {pages} {013009} (\bibinfo {year} {2005})},\ \Eprint
  {http://arxiv.org/abs/hep-ph/0503283} {arXiv:hep-ph/0503283 [hep-ph]}
  \BibitemShut {NoStop}%
\bibitem [{\citenamefont {Raut}(2013)}]{Raut:2012dm}%
  \BibitemOpen
  \bibfield  {author} {\bibinfo {author} {\bibfnamefont {S.~K.}\ \bibnamefont
  {Raut}},\ }\href {\doibase 10.1142/S0217732313500934} {\bibfield  {journal}
  {\bibinfo  {journal} {Mod. Phys. Lett.}\ }\textbf {\bibinfo {volume} {A28}},\
  \bibinfo {pages} {1350093} (\bibinfo {year} {2013})},\ \Eprint
  {http://arxiv.org/abs/1209.5658} {arXiv:1209.5658 [hep-ph]} \BibitemShut
  {NoStop}%
\bibitem [{\citenamefont {Fukasawa}\ and\ \citenamefont
  {Yasuda}(2015)}]{Fukasawa:2015jaa}%
  \BibitemOpen
  \bibfield  {author} {\bibinfo {author} {\bibfnamefont {S.}~\bibnamefont
  {Fukasawa}}\ and\ \bibinfo {author} {\bibfnamefont {O.}~\bibnamefont
  {Yasuda}},\ }\href {\doibase 10.1155/2015/820941} {\bibfield  {journal}
  {\bibinfo  {journal} {Adv. High Energy Phys.}\ }\textbf {\bibinfo {volume}
  {2015}},\ \bibinfo {pages} {820941} (\bibinfo {year} {2015})},\ \Eprint
  {http://arxiv.org/abs/1503.08056} {arXiv:1503.08056 [hep-ph]} \BibitemShut
  {NoStop}%
\bibitem [{\citenamefont {Davidson}\ \emph {et~al.}(2003)\citenamefont
  {Davidson}, \citenamefont {Pena-Garay}, \citenamefont {Rius},\ and\
  \citenamefont {Santamaria}}]{Davidson:2003ha}%
  \BibitemOpen
  \bibfield  {author} {\bibinfo {author} {\bibfnamefont {S.}~\bibnamefont
  {Davidson}}, \bibinfo {author} {\bibfnamefont {C.}~\bibnamefont
  {Pena-Garay}}, \bibinfo {author} {\bibfnamefont {N.}~\bibnamefont {Rius}}, \
  and\ \bibinfo {author} {\bibfnamefont {A.}~\bibnamefont {Santamaria}},\
  }\href {\doibase 10.1088/1126-6708/2003/03/011} {\bibfield  {journal}
  {\bibinfo  {journal} {JHEP}\ }\textbf {\bibinfo {volume} {03}},\ \bibinfo
  {pages} {011} (\bibinfo {year} {2003})},\ \Eprint
  {http://arxiv.org/abs/hep-ph/0302093} {arXiv:hep-ph/0302093 [hep-ph]}
  \BibitemShut {NoStop}%
\bibitem [{\citenamefont {Biggio}\ \emph {et~al.}(2009)\citenamefont {Biggio},
  \citenamefont {Blennow},\ and\ \citenamefont
  {Fernandez-Martinez}}]{Biggio:2009nt}%
  \BibitemOpen
  \bibfield  {author} {\bibinfo {author} {\bibfnamefont {C.}~\bibnamefont
  {Biggio}}, \bibinfo {author} {\bibfnamefont {M.}~\bibnamefont {Blennow}}, \
  and\ \bibinfo {author} {\bibfnamefont {E.}~\bibnamefont
  {Fernandez-Martinez}},\ }\href {\doibase 10.1088/1126-6708/2009/08/090}
  {\bibfield  {journal} {\bibinfo  {journal} {JHEP}\ }\textbf {\bibinfo
  {volume} {08}},\ \bibinfo {pages} {090} (\bibinfo {year} {2009})},\ \Eprint
  {http://arxiv.org/abs/0907.0097} {arXiv:0907.0097 [hep-ph]} \BibitemShut
  {NoStop}%
\bibitem [{\citenamefont {Friedland}\ \emph {et~al.}(2004)\citenamefont
  {Friedland}, \citenamefont {Lunardini},\ and\ \citenamefont
  {Maltoni}}]{Friedland:2004ah}%
  \BibitemOpen
  \bibfield  {author} {\bibinfo {author} {\bibfnamefont {A.}~\bibnamefont
  {Friedland}}, \bibinfo {author} {\bibfnamefont {C.}~\bibnamefont
  {Lunardini}}, \ and\ \bibinfo {author} {\bibfnamefont {M.}~\bibnamefont
  {Maltoni}},\ }\href {\doibase 10.1103/PhysRevD.70.111301} {\bibfield
  {journal} {\bibinfo  {journal} {Phys. Rev.}\ }\textbf {\bibinfo {volume}
  {D70}},\ \bibinfo {pages} {111301} (\bibinfo {year} {2004})},\ \Eprint
  {http://arxiv.org/abs/hep-ph/0408264} {arXiv:hep-ph/0408264 [hep-ph]}
  \BibitemShut {NoStop}%
\bibitem [{\citenamefont {Friedland}\ and\ \citenamefont
  {Lunardini}(2005)}]{Friedland:2005vy}%
  \BibitemOpen
  \bibfield  {author} {\bibinfo {author} {\bibfnamefont {A.}~\bibnamefont
  {Friedland}}\ and\ \bibinfo {author} {\bibfnamefont {C.}~\bibnamefont
  {Lunardini}},\ }\href {\doibase 10.1103/PhysRevD.72.053009} {\bibfield
  {journal} {\bibinfo  {journal} {Phys. Rev.}\ }\textbf {\bibinfo {volume}
  {D72}},\ \bibinfo {pages} {053009} (\bibinfo {year} {2005})},\ \Eprint
  {http://arxiv.org/abs/hep-ph/0506143} {arXiv:hep-ph/0506143 [hep-ph]}
  \BibitemShut {NoStop}%
\bibitem [{\citenamefont {Oki}\ and\ \citenamefont
  {Yasuda}(2010)}]{Oki:2010uc}%
  \BibitemOpen
  \bibfield  {author} {\bibinfo {author} {\bibfnamefont {H.}~\bibnamefont
  {Oki}}\ and\ \bibinfo {author} {\bibfnamefont {O.}~\bibnamefont {Yasuda}},\
  }\href {\doibase 10.1103/PhysRevD.82.073009} {\bibfield  {journal} {\bibinfo
  {journal} {Phys. Rev.}\ }\textbf {\bibinfo {volume} {D82}},\ \bibinfo {pages}
  {073009} (\bibinfo {year} {2010})},\ \Eprint {http://arxiv.org/abs/1003.5554}
  {arXiv:1003.5554 [hep-ph]} \BibitemShut {NoStop}%
\bibitem [{\citenamefont {Fukasawa}\ and\ \citenamefont
  {Yasuda}(2016)}]{Fukasawa:2016nwn}%
  \BibitemOpen
  \bibfield  {author} {\bibinfo {author} {\bibfnamefont {S.}~\bibnamefont
  {Fukasawa}}\ and\ \bibinfo {author} {\bibfnamefont {O.}~\bibnamefont
  {Yasuda}},\ }\href@noop {} {\  (\bibinfo {year} {2016})},\ \Eprint
  {http://arxiv.org/abs/1608.05897} {arXiv:1608.05897 [hep-ph]} \BibitemShut
  {NoStop}%
\end{thebibliography}%
\end{document}